\documentclass[aps,pra,longbibliography, twocolumn,10pt,superscriptaddress]{revtex4-2}
\usepackage[utf8]{inputenc}
\usepackage[final]{graphicx}
\usepackage{times,bbm,amsmath,amssymb}
\usepackage{epsfig,color}
\usepackage{xcolor}
\usepackage{hyperref}
\hypersetup{
    colorlinks = true
}
\usepackage{cleveref}
\usepackage{microtype}

\usepackage{float}
\usepackage[caption = false]{subfig}

\usepackage[greek,english]{babel}
\usepackage{thumbpdf,enumerate}
\usepackage{booktabs}
\usepackage{sidecap}
\usepackage[scaled=.8]{couriers}
\usepackage{multirow}
\usepackage{placeins}
\usepackage{relsize}
\usepackage{pst-grad,bm}
\usepackage{epigraph}
\usepackage{longtable}
\usepackage{ulem}
\usepackage{acronym}
\usepackage{physics}
\usepackage{dsfont}

\usepackage{easyReview}
\usepackage{hyperref}

\begin{document}

\title{Variational quantum cloning machine on an integrated photonic interferometer}

\author{Francesco Hoch}
\affiliation{Dipartimento di Fisica - Sapienza Universit\`{a} di Roma, P.le Aldo Moro 5, I-00185 Roma, Italy}

\author{Giovanni Rodari}
\affiliation{Dipartimento di Fisica - Sapienza Universit\`{a} di Roma, P.le Aldo Moro 5, I-00185 Roma, Italy}

\author{Eugenio Caruccio}
\affiliation{Dipartimento di Fisica - Sapienza Universit\`{a} di Roma, P.le Aldo Moro 5, I-00185 Roma, Italy}

\author{Beatrice Polacchi}
\affiliation{Dipartimento di Fisica - Sapienza Universit\`{a} di Roma, P.le Aldo Moro 5, I-00185 Roma, Italy}

\author{Gonzalo Carvacho}
\affiliation{Dipartimento di Fisica - Sapienza Universit\`{a} di Roma, P.le Aldo Moro 5, I-00185 Roma, Italy}

\author{Taira Giordani}
\email[Corresponding author: ]{taira.giordani@uniroma1.it}
\affiliation{Dipartimento di Fisica - Sapienza Universit\`{a} di Roma, P.le Aldo Moro 5, I-00185 Roma, Italy}

\author{Mina Doosti}\affiliation{School of Informatics, University of Edinburgh, 10 Crichton Street, EH8 9AB Edinburgh, United Kingdom}

\author{Sebasti\`{a} Nicolau}
\affiliation{CNRS, LIP6, Sorbonne Université, 4 place Jussieu, 75005 Paris, France}
\affiliation{Escola T\`{e}cnica Superior d’Enginyeria de Telecomunicaci\'{o} de Barcelona, Universitat Polit\`{e}cnica de Catalunya, Carrer de Jordi Girona, 1-3, 08034 Barcelona, Spain}

\author{Ciro Pentangelo}
\affiliation{Istituto di Fotonica e Nanotecnologie, Consiglio Nazionale delle Ricerche (IFN-CNR),
Piazza Leonardo da Vinci, 32, I-20133 Milano, Italy}

\author{Simone Piacentini}
\affiliation{Istituto di Fotonica e Nanotecnologie, Consiglio Nazionale delle Ricerche (IFN-CNR), 
Piazza Leonardo da Vinci, 32, I-20133 Milano, Italy}

\author{Andrea Crespi}
\affiliation{Istituto di Fotonica e Nanotecnologie, Consiglio Nazionale delle Ricerche (IFN-CNR), 
Piazza Leonardo da Vinci, 32, I-20133 Milano, Italy}
\affiliation{Dipartimento di Fisica, Politecnico di Milano, Piazza Leonardo da Vinci 32, I-20133 Milano, Italy}

\author{Francesco Ceccarelli}
\affiliation{Istituto di Fotonica e Nanotecnologie, Consiglio Nazionale delle Ricerche (IFN-CNR), 
Piazza Leonardo da Vinci, 32, I-20133 Milano, Italy}

\author{Roberto Osellame}
\affiliation{Istituto di Fotonica e Nanotecnologie, Consiglio Nazionale delle Ricerche (IFN-CNR), 
Piazza Leonardo da Vinci, 32, I-20133 Milano, Italy}

\author{Ernesto F. Galv\~{a}o}
\affiliation{International Iberian Nanotechnology Laboratory (INL), Av. Mestre José Veiga, 4715-330 Braga, Portugal}
\affiliation{Instituto de F\'isica, Universidade Federal Fluminense, Av. Gal. Milton Tavares de Souza s/n, Niter\'oi, RJ, 24210-340, Brazil}

\author{Nicol\`o Spagnolo}
\affiliation{Dipartimento di Fisica - Sapienza Universit\`{a} di Roma, P.le Aldo Moro 5, I-00185 Roma, Italy}

\author{Fabio Sciarrino}
\affiliation{Dipartimento di Fisica - Sapienza Universit\`{a} di Roma, P.le Aldo Moro 5, I-00185 Roma, Italy}

\begin{abstract}
A seminal task in quantum information theory is to realize a device able to produce copies of a generic input state with the highest possible output fidelity, thus realizing an \textit{optimal} quantum cloning machine. Recently, the concept of variational quantum cloning was introduced: a quantum machine learning algorithm through which, by exploiting a classical feedback loop informed by the output of a quantum processing unit, the system can self-learn the programming required for an optimal quantum cloning strategy. In this work, we experimentally implement a $1 \rightarrow 2$ variational cloning machine of dual-rail encoded photonic qubits, both for phase-covariant and state-dependent cloning. We exploit a fully programmable 6-mode universal integrated device and classical feedback to reach near-optimal cloning performances. Our results demonstrate the potential of programmable integrated photonic platforms for variational self-learning of quantum algorithms.
\end{abstract}

\maketitle

\section{Introduction}

In classical information theory, one can freely and perfectly measure, copy, and broadcast information without any intrinsic limitation. Conversely, one of the most distinctive features of quantum information theory is that quantum states cannot be perfectly cloned. This is a seminal result historically known as the no-cloning theorem \cite{Wootters1982, park1970concept}, and a key enabling feature of quantum cryptographic protocols \cite{Bennett2014, Aharonov2000}.
Indeed, the no-cloning theorem can be used to prove the unconditional security of quantum cryptography protocols, i.e. without relying on computational complexity results or assumptions on the resources available to an eavesdropper as in classical cryptography.
Nonetheless \textit{imperfect} cloning of quantum information is indeed possible up to fundamental upper bounds in the output state fidelity, imposed by quantum theory. The notion of \textit{optimal} cloning arose to describe cloning machines reaching the optimal possible fidelities.  This led to several proposals and implementations of such a task \cite{bruss1998optimal,buvzek1996quantum,lamas2002experimental,Ricci2004, scarani2005quantum, durt2005economical, stanev2023deterministic,zhang2012optimal,coyle2022progress,fan2014quantum,wang2011unified,xu2008experimental,liu2021all,nagali2009optimal,nagali2010experimental,niu1999two,duan1997two}.
Optimal quantum cloning procedures are useful whenever one needs to broadcast quantum information among several parties without measuring it \cite{nagali2010experimental}, {or in quantum cryptographic attacks \cite{scarani2005quantum}.}

In this context, the concept of \textit{variational quantum cloning machine} was recently introduced \cite{coyle2022progress}. {Cloning machines are an example of a genuinely quantum application in the field of quantum machine learning, where both input and output are quantum states.} This idea originated from the challenges related to non-trivial actual implementation of cloning-based attacks on cryptographic protocols \cite{scarani2005quantum}.
One challenge is the need for deep circuits that are incompatible with noisy intermediate-scale quantum (NISQ) devices \cite{preskill2018quantum}; another is the difficulty in analytically finding optimal cloning machines for particular families of states \cite{coyle2022progress}. Moreover, using a variational approach to cloning can be useful for the optimization of the protocol even when a complete characterization of a given quantum device is not experimentally feasible. Variational quantum algorithms \cite{biamonte2021universal,cerezo2021variational,wecker2015progress,biamonte2021universal,endo2021hybrid,coyle2022progress,mcclean2016theory} do indeed derive their strength from the cooperation between classical and quantum processors. In these algorithms, a quantum processor is used to estimate an objective function, while a classical computer updates the circuit parameters to optimize such a function.
Variational quantum algorithms have been used in many applications, including quantum chemistry \cite{peruzzo2014variational}, quantum algorithm discovery \cite{morales2018variational, khatri2019quantum, jones2022robust, bravo2023variational}, foundations of quantum mechanics \cite{arrasmith2019variational}, and quantum machine learning (QML) \cite{wittek2014quantum,biamonte2017quantum, kopczyk2018quantum,schuld2018supervised}.
In particular, photonic circuits represent a promising candidate platform for the implementation of QML and variational algorithms. Indeed, photons represent a naturally decoherence-free system for which single-qubit operations can be easily performed with high fidelity \cite{peters2003precise}. Furthermore, photons can be used as “flying qubits” for communication-based tasks such as networking quantum processors \cite{cacciapuoti2019quantum}. The first demonstration of a photonic variational quantum cloner is reported in \cite{javsek2019experimental}, where a single two-qubit gate with two trainable parameters was exploited to achieve optimal phase-covariant cloning by adopting a bulk optics apparatus \cite{sciarrino2007implementation,soubusta2007several,bartuuvskova2007fiber,vcernoch2006experimental,d2003optimal,chen2007experimental,fiuravsek2003optical,sciarrino2005realization,bruss2000phase,javsek2019experimental}.
However, in order to scale up the size and computational power of new photonic platforms, it is essential to investigate their general capabilities when variational approaches are used to obtain useful quantum circuit designs \cite{peruzzo2014variational,spagnolo2022experimental,pelucchi2022potential,wang2020integrated,giordani2023integrated,crespi2013integrated}.
{Furthermore, realizations of variational quantum cloning on photonic chips could enable compact,
in-line eavesdropping tests of standard quantum key distribution protocols (e.g. BB84 \cite{Bennett2014} and B92 \cite{Bennett1992}), with promising
interfaces to the existing fiber-optic and free-space communication networks \cite{Luo2023}.}

In this work, we tackle the implementation of a variational quantum cloning machine on an integrated photonic-based platform,
providing the first experimental demonstration of such an algorithm in a fully programmable integrated interferometer. By employing this platform, characterized by intrinsic stability and supporting dual-rail encoded photonic qubits, we implemented a variational cloning protocol with control of up to $12$ independent parameters. 
Our results show that we were able to train our photonic circuit to achieve near-optimal cloning fidelities both for phase-covariant \cite{sciarrino2007implementation,soubusta2007several,bartuuvskova2007fiber,vcernoch2006experimental,d2003optimal,chen2007experimental,fiuravsek2003optical,sciarrino2005realization,bruss2000phase,javsek2019experimental} and state-dependent \cite{bruss1998optimal} cloning {machines, which represent two examples of quantum eavesdropping attacks to the BB84 and B92 protocols \cite{scarani2005quantum, coyle2022progress}.} {The work proposes convenient training strategies that minimize the number of measurements for the evaluation of cost functions and enhance the trainability in multi-parameter scenarios when compared to previous implementations of variational phase-covariant cloning machines \cite{javsek2019experimental}. {By
leveraging 2-designs to reduce the number of required measurements in the optimization routine \cite{di2014short,dankert2009exact}, this work
offers an effective test-bed for the validation of variational quantum cloning protocols and benchmarking under realistic quantum cryptographic attacks.} Furthermore, we provide a systematic study of the effect of imperfect convergence of the variational algorithm on the performances of the cloning machines. Such a variational approach is very general and flexible since, leveraging on the complexity and compactness of universal and reconfigurable integrated devices, it considers as a cloning machine the most generic unitary transformation over 4 optical modes. These features allow for finding optimal cloning circuits in those cases in which an analytical solution is not trivial, as we demonstrate in the state-dependent cloning task. 
}

The paper is structured as follows. In Sec. \ref{sec:theo}, we introduce the theoretical framework of our work by briefly reviewing the different types of cloning machines, while in Sec. \ref{sec:implementation}, we present and describe our experimental apparatus.
Then, the experimental results regarding the implementation of phase-covariant and state-dependent cloning protocols are described respectively in Sec. \ref{sec:phase_covariant} and in Sec. \ref{sec:state_dependent}. Finally, we provide a summary of the work and an overview of future perspectives.

\begin{figure}[t]
    \centering
    \includegraphics[width = \columnwidth]{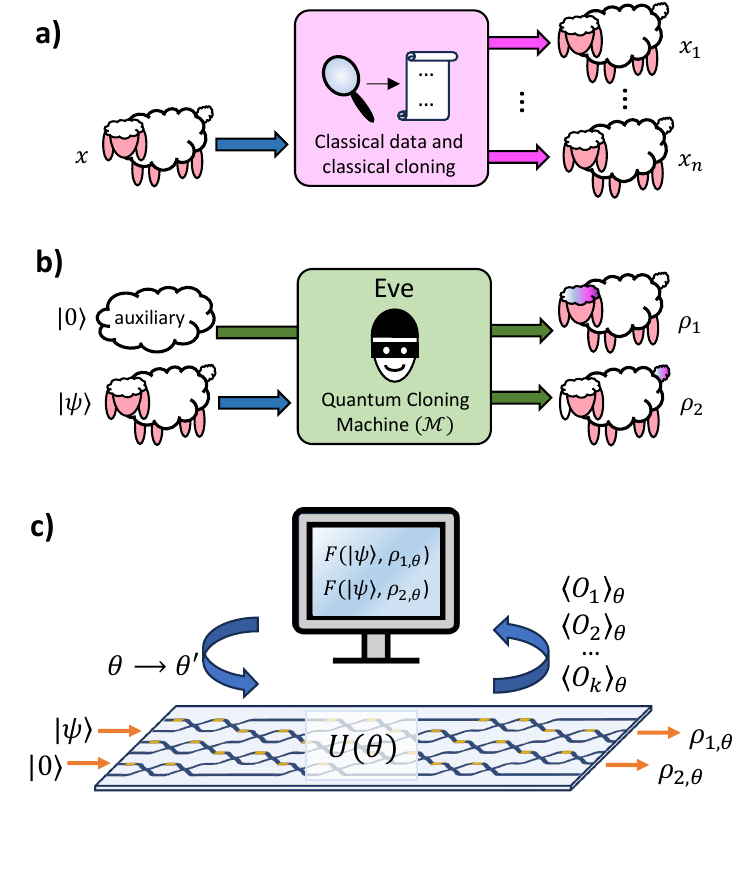}
    \caption{\textbf{Classical and quantum cloning process. a)} In a classical scenario it is always possible to read a variable $x$ and create as many copies as needed. \textbf{b)} In a quantum context, it is not possible to recover exactly all the properties of a generic state using a finite set of identical copies. This implies that any quantum cloning protocol will generate errors during the process. \textbf{c)} A photonic integrated circuit and a classical computer can work in synergy to implement a variational algorithm for the implementation of the cloning process. One qubit is prepared in the state to be cloned, $\ket{\psi}$, and another in an auxiliary initial state $\ket{0}$. At each run, the two qubits interfere in a photonic circuit whose parameters are updated according to the output states $\rho_{1,\theta}$ and $\rho_{2,\theta}$. The circuit is trained according to an appropriate loss function such that the fidelity between the two output qubits and $\ket{\psi}$ is maximized for both copies.}
    \label{fig:concept}
\end{figure}

\section{Theoretical framework}
\label{sec:theo}

In classical information theory, data is stored in elementary units of information known as bits, which can be read and copied deterministically without any \textit{a priori} loss of information, as illustrated in Fig.~\ref{fig:concept}a. In quantum information theory, conversely, information can be stored in a state $\ket{\psi}$ featuring quantum superposition. The no-cloning theorem \cite{Wootters1982} states that the task of perfectly cloning a quantum state -- in a deterministic fashion, with unit fidelity, and for every arbitrary input state -- is forbidden by quantum mechanics. Indeed, due to the postulates of quantum mechanics, extracting any information from a single copy of a quantum state by applying projective measurements destroys its superposition state. A consequence of this is the impossibility of performing a unitary transformation able to produce two perfect copies of an arbitrary input quantum state. Formally, quantum mechanics forbids the following process:
\begin{equation}
    U\ket{\psi}_1 \ket{0}_2\ket{\mathcal{M}}_\mathcal{M} \; \rightarrow \; \ket{\psi}_1\ket{\psi}_2\ket{\mathcal{M}'}_\mathcal{M} \quad \forall \ket{\psi}.
\end{equation}
Here, the first qubit $\ket{\psi} = \cos(\theta) \ket{0} + \sin(\theta) e^{i\phi} \ket{1}$ is an arbitrary normalized two-dimensional state, the second qubit used to write the state to be cloned is initialized in state $\ket{0}$, and $\ket{\mathcal{M}}_\mathcal{M}$ and $\ket{\mathcal{M}'}_\mathcal{M}$ are ancillary states describing the internal state of the cloning machine.

While perfect cloning of quantum information is not possible, the seminal result of Ref. \cite{buvzek1996quantum} showed that it is possible to devise a machine able to perform a process that obtains imperfect clones with a certain average fidelity strictly smaller than one, which can be optimised to an \textit{optimal} cloner; i.e. achieving the upper fidelity bound allowed by quantum mechanics, as depicted in  Fig.~\ref{fig:concept}b. %
Indeed, in the so-called $1 \rightarrow 2$ scenario studied in \cite{buvzek1996quantum}, the authors discussed a unitary operation $U$ realizing the following transformation:
\begin{equation}
    U \ket{\psi}_1 \ket{0}_2 \ket{\mathcal{M}}_\mathcal{M} \; \rightarrow \; \ket{\Psi}_{12\mathcal{M}},
\end{equation}
where the indices $\{1,2,\mathcal{M}\}$ identify the Hilbert spaces where the two output states and the state of the machine lie, respectively. The state $\ket{\psi}_1$ to be cloned is mapped into two output states, according to output qubit states $1$ and $2$, described respectively by the reduced density matrices:
\begin{equation}
    \rho_1 = \rho_2 = \frac{5}{6} \dyad{\psi} + \frac{1}{6} \dyad{\psi_\bot}.
\end{equation}
This means that the output quantum state fidelity with the original input state $\sigma = \dyad{\psi}$ is found to be:
\begin{equation}
    F_{1,2} = \Tr(\sqrt{\sqrt{\sigma} \rho_{1,2} \sqrt{\sigma}})^2  = 5/6,
\end{equation}
which coincides with the upper bound of the cloning fidelity for the $1 \rightarrow 2$ scenario with qubits.

Following this result, several variants of imperfect cloning were proposed in the literature, whose features depend on the requirements and the assumptions being made on the cloning process \cite{bruss1998optimal,buvzek1996quantum,lamas2002experimental,stanev2023deterministic,zhang2012optimal,coyle2022progress,fan2014quantum,wang2011unified,xu2008experimental,liu2021all,nagali2009optimal,nagali2010experimental,scarani2005quantum,durt2005economical,niu1999two,duan1997two,duan1998probabilistic, Ricci2004}. Typically, those variants have been classified depending on the properties we now list below:

\textit{Dimensionality of the input states. --} Cloning machines can be devised to clone states in any Hilbert space dimension. The cloning of qubits, i.e. two-dimensional quantum states, has been widely investigated. Moreover, there have been works covering 
the cloning of qutrits \cite{Cerf2002}, qudits \cite{Fan2003} or continuous variables systems \cite{Cerf2002a}.

\textit{Number of input/output copies. --} A second classification is based on the number $N$ of copies of a state submitted to the cloner, and the number $M$ of copies provided by the cloner, usually indicated as a $N\rightarrow M$ cloning machine. The most commonly studied type of cloner is the $1\rightarrow 2$ cloner, but generalizations to an arbitrary number of input/output copies $N\rightarrow M$ have been investigated \cite{Zhang2010}.

\textit{Dimension of the auxiliary states. --} In a typical scenario, one does not place limitations on the computing power of a cloning machine, given the interest in studying the ultimate quantum limit. However, from a practical point of view, it can be relevant to identify the limits of a quantum cloning process when limitations are placed on the resources to be used. In particular, limiting the size of the auxiliary space used has led to the concept of \textit{economical quantum cloning}  \cite{durt2005economical, niu1999two}.

\textit{Ensemble of input states. --} This classification is based on the ensemble of states to be given to the cloning machine. For instance, one may require the cloning machine to copy unknown states belonging to specific sets of states in a given dimension. In particular, for the qubit cloning machines, one can identify three main relevant classes: (i) if the set of states to be cloned are drawn from the uniform, Haar distribution from the whole Bloch sphere, the cloner is called \textit{universal}, (ii) if we restrict the set of states to those uniformly picked from a great circle of the Bloch sphere, then the machine is called a \textit{phase-covariant cloner} \cite{Fan2001}, and (iii) if the set is further restricted to just two states, not necessarily orthogonal, the machine is a \textit{state-dependent} cloner \cite{bruss1998optimal}.

\textit{Symmetric/asymmetric cloner. --} Instead of producing several quantum clones with the same fidelity, in some situations it can be useful to consider unbalanced cloners whose output clones have different fidelities. A cloner is said to be \textit{symmetric} if all copies are equivalent and have the same fidelity with the input state. Conversely, a clone is said to be \textit{asymmetric} \cite{Hashagen2017} when the copies do not have the same fidelity with respect to the input state.

\section{Photonic implementation of a variational cloning machine}
\label{sec:implementation}

\begin{figure*}[ht]
    \centering
\includegraphics[width=\textwidth]{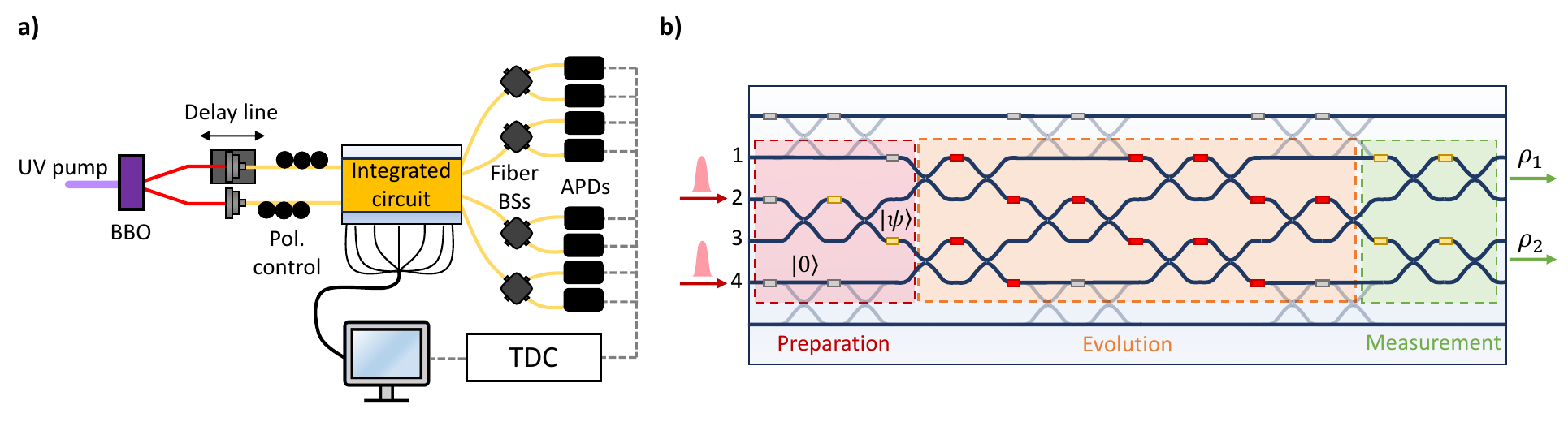}
    \caption{\textbf{Experimental implementation of the Variational Cloning Machine. a)} Photons are generated through spontaneous parametric down-conversion via a $392.5~$nm UV pulsed laser pumping a Beta-Barium Borate (BBO) crystal to generate $785~$nm photon pairs. To ensure indistinguishability among such photons, temporal synchronization is achieved through a free space delay line. The photons are then coupled to single-mode fibers, while polarization is controlled through in-fiber paddles. Photons are then injected into a six-mode universal photonic integrated circuit. At the output of the integrated device, we collect the four central output modes and resolve the number of photons via fibered beam splitters (BSs), eight avalanche photodiodes (APDs), and a time-to-digital converter (TDC), thus implementing a pseudo-photon-number resolving detection scheme. The output probability distribution is fed to a classical computer in order to estimate the fidelity between the output clones and the input target state. This allows the computation of the cost function to be minimized via an update of the circuit parameters.
    \textbf{b)} Detail of the programmable integrated photonic circuit, which can be divided into three stages: preparation of the input state to be cloned, implementation of the variational cloning machine, and measurement of the output state fidelity. In the red preparation stage, two initial qubits are prepared respectively in states $\ket{0}$, dual-rail encoded in modes 1 and 4, and $\ket{\psi}$, in modes 2 and 3. Then, in the orange region, a 4-mode universal interferometric mesh is set up to be tuned by the classical algorithm. The red rectangles represent the 12 internal phases optimized in the process, while the dark blue crossings are fixed unbiased integrated beamsplitters. Finally, the green area is configured so as to project the two clones onto the target state for an estimation of the quantum state fidelity. The grey phase shifters remain unused; the yellow ones are used for preparation and measurement, and the red ones are tuned variationally.
    At each run of the algorithm, the fidelity between the output states $\rho_{1}$ and $\rho_{2}$ and the target input state $\ket{\psi}$ is evaluated and used to update the variational parameters, via the Nelder-Mead classical optimization algorithm.}
    \label{fig:vqa_scheme}
\end{figure*}

Let us now describe our experimental apparatus for the photonic implementation of a variational quantum cloning machine, which is depicted in Fig.~\ref{fig:vqa_scheme}a.
Initially, pairs of single photons at $785$~nm are generated via type-II spontaneous parametric down-conversion achieved in a beta-barium borate (BBO) crystal pumped with a $392.5$~nm UV pulsed laser, with a repetition rate of $76$~MHz. The two emitted photons are spectrally filtered through a $3~$nm band-pass filter, while their polarization is fixed via two polarizing beam splitters before being coupled into single-mode fibers. To guarantee pairwise indistinguishability between the generated single photons, they are temporally synchronized by tuning their arrival time via a free-space delay line placed on one of the optical fibers. Polarization indistinguishability is optimized through two in-fiber paddle controllers.
Next, the photons are injected into a six-mode fully programmable photonic integrated circuit
\cite{Pentangelo2024,giordani2023experimental}
through a fiber array which is directly attached to the chip.
The circuit, fabricated by femtosecond laser waveguide writing \cite{Corrielli2021, Ceccarelli2020}, and shown in Fig.~\ref{fig:vqa_scheme}b, consists of a set of directional couplers and thermo-optical phase shifters arranged according to the scheme reported in Ref. \cite{Clements2016}, enabling the implementation of any possible unitary evolution between input and output modes. We note that the reconfigurability of the interferometer transformation is provided via $30$ thermo-optical resistive phase shifters controlled in current \cite{flamini2015thermally}.
After their evolution, the photons exiting from the employed modes are collected by a second fiber array, and subsequently routed to four multimode in-fiber beam splitters and eight avalanche photodiodes. In such a way, it is possible to probabilistically discriminate one and two-photon events in a single output of the integrated interferometer, thus implementing pseudo-photon-number-resolving detection (more details can be found in the Supplementary Materials). 
A preliminary calibration procedure of the programmable circuit is performed with classical laser light to reconstruct the relationship between the currents applied to the thermal shifters and the internal phases locally induced on the waveguides.
Starting from the knowledge of the calibration parameters 
we implemented a Python class that allows us to easily invert the control problem.
More specifically, given an internal configuration of the interferometer to be implemented, the developed software computes the appropriate currents and drives the power supply connected to the interferometer, enabling full control of the device from a classical computer. Further details on calibration and control of the device are reported in the Supplementary Materials.
To implement variational quantum cloning, we configure the photonic chip as shown in Fig.~\ref{fig:vqa_scheme}b. The first part, enclosed in the red area, performs the preparation of the two qubits.
In detail, we only use the four central modes of the six-mode universal integrated interferometer by excluding external modes $0$ and $5$, while photon pairs are injected in modes $2$ and $4$. In such a way, the input qubit to be cloned $\ket{\psi}$ is dual-rail encoded in modes $2$ and $3$ via a tunable beamsplitter and a phase shifter, which enables the mapping $\ket{0} \rightarrow \alpha \ket{0} + \beta e^{i\phi} \ket{1}$, where $\alpha$ and $\beta$ are real-valued parameters. Conversely, the auxiliary qubit on which we wish to write a copy of the state $\ket{\psi}$ is initialized in the dual-rail state $\ket{0}$ on modes $1$ and $4$. The injected photons then undergo the transformation yielded
by the variational circuit shown in the orange area of Fig. 2b. We consider as a variational circuit a 4-mode universal interferometer that has the minimal layout in terms of depth and number of optical components \cite{Clements2016} to implement any generic linear optical transformation for different quantum cloning tasks. Such a circuit with minimal layout can be controlled via 12 phases.
In each iteration, the 12 phases depicted in red in Fig.~\ref{fig:vqa_scheme}b are
updated via the variational algorithm, according to a classical measurement-feedback informed optimization loop.
This involves performing projective measurements after the state evolution, to evaluate the fidelity between each output clone and the input state $\ket{\psi}$.
This is done by encoding the projector on the basis containing the pure input state in the last section of the chip, depicted in green in Fig.~\ref{fig:vqa_scheme}b (see also Supplementary Materials).
Such a procedure for the fidelities estimation reduces the number of measurements that a full qubit tomography would require. It fits very well with the need for a variational algorithm to estimate the cost functions on a training set iteratively, thus reducing the measurement time required for a single algorithm iteration. In addition, optimizations based on state fidelity estimations via direct projections onto the target states are more convenient than the computation of the overall fidelity with the optimal cloning circuit that needs a larger set of measurements. Furthermore, we note that the optimal cloning circuit might not be unique or known a priori.

To carry out the measurements, we consider the first copy of the initial state to be encoded in the output modes $1$ and $2$, and the second copy in modes $3$ and $4$. The data are collected by post-selecting events with a two-fold coincidence detection between one photon exiting one of the output modes 1-2, and the other exiting one of the output modes 3-4.

Given the classification provided in the previous section, our device enabled us to implement two different types of symmetric $1\rightarrow 2$ variational qubit cloners, i.e. the phase-covariant and state-dependent cloners. Our cloning machine works in a post-selected regime since it is implemented via linear optics and, given that there are no ancillary modes nor photons, it can be classified as an economical cloner.

\section{Symmetric phase-covariant cloning}
\label{sec:phase_covariant}

\begin{figure*}[t]
    \centering
    \includegraphics[width = \textwidth]{cloning_CV_legend.pdf}
    \caption{\textbf{Experimental results on phase-covariant cloning.} \textbf{a-c)} Values, as a function of the Nelder-Mead algorithm iterations, of the clones' fidelities $F_1$ and $F_2$ w.r.t. the input quantum state $\ket{\psi_\theta}$, and of the cost function $\mathcal{C}_{PC}(\vec{\theta})$ respectively. In lighter colours, we represent the values attained by the respective quantities on each point in the 12-dimensional parameter space explored by the Nelder-Mead algorithm. In darker colours, we depict the optimal result for each quantity found by the algorithm up to a given iteration. The red dashed vertical lines correspond to the iterations where the minimization algorithm was rebooted, to prevent the algorithm from getting stuck in local minima due to the presence of experimental noise and fluctuations related to the finite measurement statistics. 
    The yellow dashed line represents the average theoretical optimal value 
    of the variational protocol. \textbf{d)} Bloch sphere with the four states used for the training of the algorithm (in the blue colors); the optimal circuit was tested on a larger number of states drawn from the equator of the sphere (purple circle). \textbf{e)} Validation of the optimal cloning circuit found by the variational protocol. The output fidelity $F_{\mathrm{test}}$ of the two clones ($F_1$ and $F_2$) is evaluated for 50 test input states distributed evenly on the Bloch sphere equator. Here, statistical errors are estimated from a Poisson distribution of the two-fold coincidence counts probed at the output of the measurement stage. The yellow dashed line reports the theoretical mean fidelity for a symmetric phase-covariant cloner ($F_O = 0.853$). The green dashed line represents the theoretical upper limit on the average output fidelity of a semi-classical measure-and-prepare approach.}
    \label{fig:data_PC}
\end{figure*}

The first variant of variational cloning machine that we aim to implement experimentally is a \textit{symmetric phase-covariant cloner} (PC), which aims to optimally clone states that lie on the equatorial plane of the Bloch sphere $ \ket{\psi_\phi} = \frac{\ket{0}+e^{i \phi} \ket{1}}{\sqrt{2}}$, with the phase parameter $\phi$ taking arbitrary values.
This machine represents a useful primitive for cryptographic attacks on the BB84 protocol and on the coin-tossing protocol \cite{coyle2022progress}. 
To implement such a cloning machine with a variational approach, one needs to choose an appropriate cost function $\mathcal{C}$. Such a function, on one side, must depend on the set of parameters $\vec{\theta}$ defining the evolution implemented on the quantum device while, on the other, has to be computed from the statistics obtained at the output of the quantum processor. In such a way, one can employ a classical optimization algorithm to iteratively search the 12-dimensional parameter space and find the minima of the cost function, thus searching $\min_{\vec{\theta}} \mathcal{C}(\vec{\theta})$ with $ \vec{\theta} \in [0,2\pi]^{12}$. If the cost function has been chosen properly, this corresponds to the optimal circuit configuration to solve the task at hand.

It has been shown \cite{coyle2022progress} that the following cost function captures the aforementioned requirements:
\begin{equation}
    \mathcal{C}_\textbf{PC} = {\mathds{E}} \biggl[ (1-F_{1,\phi})^2+(1-F_{2,\phi})^2 +(F_{1,\phi}-F_{2,\phi})^2\biggr],
    \label{eq:cost_function_ph}
\end{equation}
where with $F_{1,\phi}$ and $F_{2,\phi}$ we indicate the output \textit{local} fidelities of the two copies with a given input state $\ket{\psi_\phi}$, defined as:
\begin{equation}
    F_{i,\phi} = \Tr(\sqrt{\sqrt{\sigma_\phi} \rho_i \sqrt{\sigma_\phi}})^2.
\end{equation}
with $\sigma_\phi = \dyad{\psi_\phi}$. In Eq. \eqref{eq:cost_function_ph}, ${\mathds{E}}[\cdot]$ stands for the expected value of the quantity inside the brackets over all possible input states, i.e. averaged over all angles $\phi \in [0,2\pi)$. Note that the first two terms of the cost function push the algorithm to look for the optimal output fidelities of both copies, while the third term enforces the symmetry of the two output fidelities.

In a practical scenario, it is experimentally demanding to directly probe the cost function in Eq.~\eqref{eq:cost_function_ph}, since it requires computing an average over a continuum of states.
To circumvent this issue, we evaluate the cost function over a small finite set of states for which the expectation value in Eq.~\eqref{eq:cost_function_ph} over the set is equivalent to the continuous expectation value over the equatorial X-Y plane, where the phase-covariant states lie. To optimise the efficiency of the evaluation, we choose a minimal set for which the above property holds. This is achieved using prior information of the cloning ensemble, and properties of designs~\cite{di2014short}. 
As such, the approximated cost function at each iteration is:
\begin{equation}
    \mathcal{C}_\textbf{PC} = \sum_{\phi \in \mathcal{F}} (1-F_{1,\phi})^2+(1-F_{2,\phi})^2 +(F_{1,\phi}-F_{2,\phi})^2,
    \label{eq:cost_tdes}
\end{equation}
where the set of input states employed to evaluate the cost function corresponds to the phases $\mathcal{F} = [0,\frac{\pi}{2}, \pi, \frac{3\pi}{2}]$. Such phases define the set of states composed by $\left(\ket{0} \pm \ket{1}\right)/\sqrt{2}$ and $\left(\ket{0} \pm i \ket{1}\right)/\sqrt{2}$, i.e., eigenstates of Pauli's operators $X$ and $Y$.

We now show why the above cost function is a valid approximation and the set of chosen states is a minimal set for our problem. First, we observe that the cost function is of the form $(1 - F_{\phi})^2$ where $F_{\phi}$ is the fidelity of the cloned state with the pure target state, which is always on the X-Y plane for this phase-covariant cloning. Given that the target state is pure, and the clone is a general mixed state $\rho$, the fidelity is given by $\bra{\psi_{\phi}}\rho \ket{\psi_{\phi}} = \mathrm{Tr}[\dyad{\psi_\phi} \rho]$. So the cost function includes terms of at most second-moment (due to the $F^2$ term).
{In this case, one can approximate an average over the uniform measure of the space of interest, with a 2-design \cite{dankert2009exact,di2014short} over the same space, which is a finite set of pure states whose first and second statistical moments match those of the uniform distribution.}
Interestingly, here we do not need to use a state 2-design (i.e. a spherical design) because the states lie on the equatorial plane only. One can construct a design for that plane by only using the 4 points on the circle, corresponding to the eigenstates of Pauli $X$ and $Y$. We also emphasise that the set is minimal as this is the minimum number of states for which one can achieve an exact planar 2-design, and hence one can easily show that
\begin{equation}
    \frac{1}{2\pi} \int (1 - F_{\phi})^2 d\phi = \frac{1}{4} \sum_{i \in \mathcal{F}} (1 - F_i)^2.
\end{equation}
Note that we can look at the form of each term in the cost function of Eq.~\eqref{eq:cost_function_ph} separately since we are interested in the symmetric cloning machine.
{This approach presents an exact estimation with a limited number of measurements, $4$ in our case. The uncertainty in such an estimate will depend only on the experimental errors in the detection stage, such as Poissonian fluctuations in single-photon counts or detector dark counts. On the contrary, the common approach that uses the average over a set of $n$ random points on the equator requires balancing between the number of measurements $n$ and the intrinsic additional approximation error $\epsilon$ of such a technique that scales as $O(1/\sqrt{n})$.
}
We note that the method can be used as a general strategy to enhance training efficiency and can be extended to any scenario and dimension where there is prior information about the subspace spanned by the training states. As long as a uniform measure can be defined over the subspace, the cost function only needs to be evaluated using a $t$-design (where $t$ matches the moment of the cost function) over that subspace.

Due to the complex landscape of the cost function to be optimized, we employ the Nelder-Mead (NM) optimization algorithm \cite{nelder1965simplex}. This method is a well-known simplex-based, gradient-free optimization protocol which was already employed in a photonic-based setting \cite{poderini2022ab, javsek2019experimental}. In brief, NM is able to explore the $d$-dimensional parameter space starting from an initial set of $d+1$ points $\Theta_0 = \{\vec{\theta}_0,\dots,\vec{\theta}_{d+1}\}$ forming a simplex, on which the cost function is initially evaluated.  At each algorithm iteration, by computing again the cost function in another point chosen according to a set of fixed geometrical rules, the simplex is updated in such a way that it progressively converges around a minimum of the given cost function. By applying the NM algorithm to the set of phases defining the evolution programmed in the photonic integrated device, with the goal of minimizing the cost function of Eq.~\eqref{eq:cost_tdes}, we are able to converge to an optimal symmetric phase-covariant process via a variational approach, as shown in Fig.~\ref{fig:data_PC}a-c. The states used for the training phase are instead reported on the Bloch sphere in Fig.~\ref{fig:data_PC}d.

In Fig.~\ref{fig:data_PC}a-c, we show the evolution during the NM algorithm optimization of the three relevant figures of merit of the variational cloning machine, namely the output state fidelities $F_1$, $F_2$ and the cost function $\mathcal{C}_{PC}(\vec{\theta}_i)$. Specifically, we show both the values attained by each quantity in each configuration explored by the minimization process and the optimal values of each quantity found by the algorithm up to the considered iteration. We note that the only quantity that is tracked by the classical optimization routine is the cost function $\mathcal{C}_{PC}(\vec{\theta}_i)$. Due to the presence of experimental noise and fluctuation, the NM algorithm can get stuck in a point in the parameter space which is not an actual minimum of the cost function, a known issue of the NM algorithm related to the premature \textit{collapse} of the simplex in an unwanted point. Within the present protocol, we circumvent this issue by implementing a \textit{reboot} procedure \cite{Huang2018}, in which the algorithm has been manually restarted by reinitializing an enlarged simplex, at the iterations indicated by the red lines in the figure. The yellow dashed lines show the theoretical limit for the cost function and the corresponding output state fidelities. We observe that the minimization approaches the theoretical limit relatively well but does not completely reach the optimal value, mainly due to the presence of experimental noise.

After $\sim 700$ iterations, we then halt the minimization process and proceed with the validation of the optimal circuit found by the variational process. To do so, we evaluate the quantum state fidelity of each of the two output clones for $50$ different input states evenly distributed at intervals of $2\pi/50$ radians on the equator of the Bloch sphere. The results of such a validation stage, depicted in Fig.~\ref{fig:data_PC}e,
can be compared against the fidelity achievable by a semi-classical measure and prepare scenario, and with an ideal optimal symmetric cloning machine. The measure-and-prepare setting involves a measurement along a random basis on the equator of the Bloch sphere, with the measurement outcome used to prepare two imperfect copies of the original state, resulting in an average fidelity w.r.t. the original state of $F_{SC} = 0.750$. The optimal average cloning fidelity corresponds to $F_O = 0.853$ \cite{Fan2001}.

With our variational cloning machine, we reached an average output fidelity of $F_1^{opt} = 0.802 \pm 0.001$ and $F_2^{opt} = 0.838 \pm 0.001$. The values of the fidelities achieved for each of the 50 states are all at least three standard deviations above the semi-classical limit, showcasing the feasibility of the variational approach for the implementation of a phase-covariant quantum cloner. We note that two averages do not go beyond the theoretical optimal limit $F_O = 0.8533$ and display a residual asymmetry. Such a discrepancy between the two cloners, the residual fringe in the phase $\phi$ on the Bloch Sphere, and the not optimal reached fidelity are compatible with an incomplete convergence of the algorithm and with the experimental noise given by the finite statistics. All these imperfections are analyzed in the Supplementary Materials.

\section {State-dependent cloning}
\label{sec:state_dependent}
\begin{figure*}[t]
    \centering
    \includegraphics[width = \textwidth]{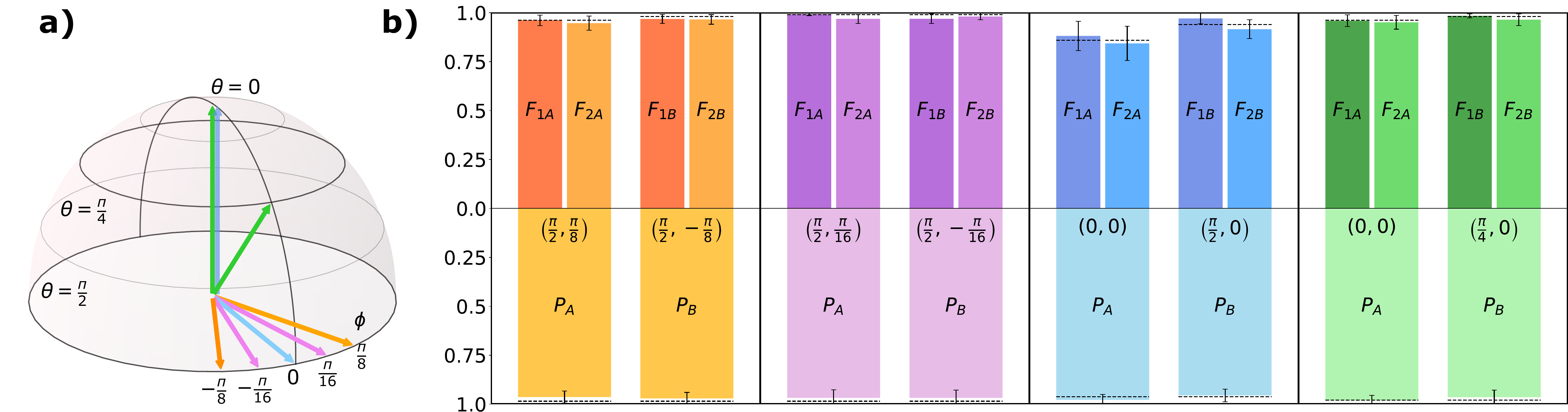}
    \caption{\textbf{Experimental results for the State-dependent variational cloning machine.} \textbf{a)} Representation on the Bloch Sphere of the qubit pairs investigated for the state-dependent cloning. Each pair is identified by a different colour. \textbf{b)} In each subplot, we report the average over four independent optimizations of output fidelities and post-selection rates for four different pairs of input states. For each pair of states $\ket{\psi_A}$ and $\ket{\psi_B}$, identified by the tuple of angles $(\theta, \phi)$ as described in the main text, we report the optimal fidelities of the two clones ($F_{1x}$ and $F_{2x}$) reached by the variational cloning machine and the corresponding post-selection probability of the cloning process ($P_x$). Here, experimental errors are computed by considering a standard Poissonian distribution over the measured two-fold coincidence events.
    With the dashed line, we report the theoretically expected optimal values computed through a numerical simulation of the variational optimization process aimed to minimize the cost function reported in Eq.~\eqref{eq:SD_CF}.}
    \label{fig:Data_SD}
\end{figure*}

To showcase the versatility of the variational approach, we have also targeted the implementation of a \textit{state-dependent} (SD) cloner.
In this variant, the set of states to be cloned is restricted to only two states $\ket{\psi_A}$ and $\ket{\psi_B}$ of the form $\ket{\psi_{A,B}}(\theta, \phi) = \cos(\theta) \ket{0} + \sin(\theta) e^{i\phi} \ket{1}$. Similarly to the previous scenario, this class of cloning machines is of interest since it can be employed to enable attacks in quantum cryptographic protocols such as the B92 one \cite{Bennett1992}, which rely on pairs of states for their correct functioning. The optimal circuit for such a cloning task is known in the qubit formalism and has been derived in Ref. \cite{bruss1998optimal}. However, the analytical derivation of the equivalent photonic circuit is challenging. This makes the variational approach the most promising to find the optimal solution.

In this scenario, we need to take into account the post-selection probability of the protocol.
Indeed, by performing numerical simulations carried out with the Quandela Perceval library \cite{heurtel2023perceval}, we found that the cost function defined in Eq.~\ref{eq:cost_function_ph} is not a suitable choice in this case since it shows the presence of numerical instabilities where the 
post-selection probability approaches values close to zero (in particular, for the optimal circuit all the fidelities are equal to one but the post-selection probability of one of the states is always equal to zero). This is reported in detail in the Supplementary Materials.
For this reason, we slightly modified the cost function by adding a regularisation term that takes into account the success rates of the protocol. Overall, the cost function in this scenario is defined as:
\begin{equation}
\begin{split}
    \mathcal{C}_{\textbf{SD}} = &(1-F_{1A})^2+(1-F_{2A})^2+(F_{1A}-F_{2A})^2+ \\
          &(1-F_{1B})^2+(1-F_{2B})^2+(F_{1B}-F_{2B})^2+ \\
          &\lambda\left[ (1-P_A)^2+(1-P_B)^2+(P_A-P_B)^2\right],
    \label{eq:SD_CF}
\end{split}
\end{equation}
where $P_A$ and $P_B$ are the post-selection probabilities for the respective states and $\lambda$ is a parameter that balances the strength of the regularisation term{. It penalizes low post-selection probabilities, preventing the
algorithm from collapsing to zero-success-probability minima,} compared to the original cost function.
{We observe robust convergence numerically for $\lambda \geq 10^{-4}$.} 
In our experiment, {we set} $\lambda = 1$ {which presents a good trade-off between fidelity optimization against success-probability preservation.
In fact, setting the $\lambda$ parameter too low considerably reduces the probability of success, which settles $P_A$ and $P_B$ around values of $0.2-0.3$, without a real gain in fidelity, while setting the parameter too large increases the probability of success at the expense of a significant reduction in fidelity.} 
In Fig.~\ref{fig:Data_SD}, we show the experimental results in terms of fidelities and post-selection probability achieved over four optimization runs for different pairs of states.
Statistical errors are computed by considering that the two-fold coincidences are distributed according to a Poissonian distribution. In the figure, we report the theoretically achievable optimal results retrieved via numerical simulations of the optimization process. In all cases, the experimentally implemented variational cloning machines are able to achieve an output fidelity and 
post-selection probability within one standard deviation of the expected value, showing again that this approach is well suited for the implementation of a state-dependent cloner, independently from the choice of the state pair to be cloned. Additionally, it can be observed that the measured fidelity values are higher than the average fidelities of the phase-covariant cloner. This behaviour is expected, since restricting the set of admissible states provides more information to the cloning machine, which can, as a result, reach higher fidelities.

\section{Discussion}
In this work, we presented and validated the first experimental implementation of a variational cloning machine on a photonic platform based on a universal programmable integrated interferometer. More specifically, we targeted the implementation of a class of $1 \rightarrow 2$ post-selected economical cloning machines of dual-rail encoded photonic qubits manipulated in a fully programmable 6-mode photonic chip. We approached the problem by employing a classical feedback-loop routine to tune the $12$ parameters that individuate a 4-mode universal linear optical interferometer in order to find the optimal cloning circuit for the specific task at hand. Such a layout allows us to tackle the problem of variational cloning even in cases where an analytical solution for a photonic linear optical circuit is not trivial, as is the case for the state-dependent machine. This variational search is performed via the Nelder–Mead algorithm, i.e. simplex-based gradient-free optimisation algorithms, and by engineering a suitable cost function such that its minima correspond to the optimal circuit state.  We experimentally prove the feasibility of the approach by implementing, at first, an optimal symmetric phase-covariant cloner, achieving an almost optimal circuit in under 1000 iterations and testing the resulting cloning machine over a set of $50$ states evenly distributed over the Bloch sphere equator. This shows that the performances of the cloning machine are near the optimal theoretical values, and are better than the best semi-classical strategy. Owing to the versatility of the variational approach paired with a programmable photonic integrated device, we were also able to implement a state-dependent cloner, successfully performing the training over different pairs of input states and obtaining results compatible with the optimal theoretical predictions estimated via numerical simulations.

In such a way we provided an experimental test of the potential of variational routines applied to a hybrid photonic platform. This work proposes effective strategies for training cloning machines by leveraging computational fidelities with minimal and optimal sets of target states. Specifically, we used as target states those commonly employed in cryptographic protocols such as the encoding and decoding bases in quantum key distribution protocols like BB84 \cite{Bennett2014} and B92 \cite{Bennett1992}. The knowledge of these bases is supposed to be available to eavesdroppers since it does not undermine the security of the shared key \cite{Bennett2014}, thus making this variational cloning paradigm an effective tool for quantum cryptographic attacks. 
In this context, the method demonstrates efficient optimization of cost functions even for the challenging scenario of state-dependent cloning machines where analytical solutions for a photonic circuit are not straightforward, which allowed us to achieve near unity state fidelities and success probabilities. The development of these variational optimizations over a photonic encoded platform makes the approach feasible to be interfaced with the current quantum-secured networks infrastructure that mainly employed photonic qubits as information carriers. In such a framework, the integrated device could, in principle, even be used to attack polarization encodings of the QKD protocols, via the use of a polarization-to-path converter prior to inputting photons onto the programmable chip \cite{Wang:16}. Moreover, the work proposes a training strategy inspired by the t-designs \cite{Ambainis2007, dankert2009exact,di2014short} that, making use of limited knowledge about the structure of the training set, significantly decreases the number of measurements for the estimations of cost functions. At the same time, it enhances the effectiveness in training multi-parameter variational circuits, paving the way to applications in the more general framework of self-learning algorithms, in which the quantum hardware learns how to optimize its internal parameters.  Further improvements of the protocol aiming at enhancing its scalability could be the inclusion of other terms in the cost function which take into account the overall depth of the circuit, towards the addition of a structural learning step in the optimization. In this way such \textit{self-learning} approaches could prove useful for near-term implementations of complex photonic quantum algorithms even when a direct mapping, from algorithm to its implementation on the quantum device, is not straightforward, paving the way for the application of those technologies in novel applications of quantum technologies.



\begin{thebibliography}{77}%
\makeatletter
\providecommand \@ifxundefined [1]{%
 \@ifx{#1\undefined}
}%
\providecommand \@ifnum [1]{%
 \ifnum #1\expandafter \@firstoftwo
 \else \expandafter \@secondoftwo
 \fi
}%
\providecommand \@ifx [1]{%
 \ifx #1\expandafter \@firstoftwo
 \else \expandafter \@secondoftwo
 \fi
}%
\providecommand \natexlab [1]{#1}%
\providecommand \enquote  [1]{``#1''}%
\providecommand \bibnamefont  [1]{#1}%
\providecommand \bibfnamefont [1]{#1}%
\providecommand \citenamefont [1]{#1}%
\providecommand \href@noop [0]{\@secondoftwo}%
\providecommand \href [0]{\begingroup \@sanitize@url \@href}%
\providecommand \@href[1]{\@@startlink{#1}\@@href}%
\providecommand \@@href[1]{\endgroup#1\@@endlink}%
\providecommand \@sanitize@url [0]{\catcode `\\12\catcode `\$12\catcode
  `\&12\catcode `\#12\catcode `\^12\catcode `\_12\catcode `\%12\relax}%
\providecommand \@@startlink[1]{}%
\providecommand \@@endlink[0]{}%
\providecommand \url  [0]{\begingroup\@sanitize@url \@url }%
\providecommand \@url [1]{\endgroup\@href {#1}{\urlprefix }}%
\providecommand \urlprefix  [0]{URL }%
\providecommand \Eprint [0]{\href }%
\providecommand \doibase [0]{https://doi.org/}%
\providecommand \selectlanguage [0]{\@gobble}%
\providecommand \bibinfo  [0]{\@secondoftwo}%
\providecommand \bibfield  [0]{\@secondoftwo}%
\providecommand \translation [1]{[#1]}%
\providecommand \BibitemOpen [0]{}%
\providecommand \bibitemStop [0]{}%
\providecommand \bibitemNoStop [0]{.\EOS\space}%
\providecommand \EOS [0]{\spacefactor3000\relax}%
\providecommand \BibitemShut  [1]{\csname bibitem#1\endcsname}%
\let\auto@bib@innerbib\@empty
\bibitem [{\citenamefont {Wootters}\ and\ \citenamefont
  {Zurek}(1982)}]{Wootters1982}%
  \BibitemOpen
  \bibfield  {author} {\bibinfo {author} {\bibfnamefont {W.~K.}\ \bibnamefont
  {Wootters}}\ and\ \bibinfo {author} {\bibfnamefont {W.~H.}\ \bibnamefont
  {Zurek}},\ }\bibfield  {title} {\bibinfo {title} {A single quantum cannot be
  cloned},\ }\href {https://doi.org/10.1038/299802a0} {\bibfield  {journal}
  {\bibinfo  {journal} {Nature}\ }\textbf {\bibinfo {volume} {299}},\ \bibinfo
  {pages} {802–803} (\bibinfo {year} {1982})}\BibitemShut {NoStop}%
\bibitem [{\citenamefont {Park}(1970)}]{park1970concept}%
  \BibitemOpen
  \bibfield  {author} {\bibinfo {author} {\bibfnamefont {J.~L.}\ \bibnamefont
  {Park}},\ }\bibfield  {title} {\bibinfo {title} {The concept of transition in
  quantum mechanics},\ }\href {https://doi.org/10.1007/bf00708652} {\bibfield
  {journal} {\bibinfo  {journal} {Foundations of Physics}\ }\textbf {\bibinfo
  {volume} {1}},\ \bibinfo {pages} {23–33} (\bibinfo {year}
  {1970})}\BibitemShut {NoStop}%
\bibitem [{\citenamefont {Bennett}\ and\ \citenamefont
  {Brassard}(2014)}]{Bennett2014}%
  \BibitemOpen
  \bibfield  {author} {\bibinfo {author} {\bibfnamefont {C.~H.}\ \bibnamefont
  {Bennett}}\ and\ \bibinfo {author} {\bibfnamefont {G.}~\bibnamefont
  {Brassard}},\ }\bibfield  {title} {\bibinfo {title} {Quantum cryptography:
  Public key distribution and coin tossing},\ }\href
  {https://doi.org/10.1016/j.tcs.2014.05.025} {\bibfield  {journal} {\bibinfo
  {journal} {Theoretical Computer Science}\ }\textbf {\bibinfo {volume}
  {560}},\ \bibinfo {pages} {7–11} (\bibinfo {year} {2014})}\BibitemShut
  {NoStop}%
\bibitem [{\citenamefont {Aharonov}\ \emph {et~al.}(2000)\citenamefont
  {Aharonov}, \citenamefont {Ta-Shma}, \citenamefont {Vazirani},\ and\
  \citenamefont {Yao}}]{Aharonov2000}%
  \BibitemOpen
  \bibfield  {author} {\bibinfo {author} {\bibfnamefont {D.}~\bibnamefont
  {Aharonov}}, \bibinfo {author} {\bibfnamefont {A.}~\bibnamefont {Ta-Shma}},
  \bibinfo {author} {\bibfnamefont {U.~V.}\ \bibnamefont {Vazirani}},\ and\
  \bibinfo {author} {\bibfnamefont {A.~C.}\ \bibnamefont {Yao}},\ }\bibfield
  {title} {\bibinfo {title} {Quantum bit escrow},\ }in\ \href
  {https://doi.org/10.1145/335305.335404} {\emph {\bibinfo {booktitle}
  {Proceedings of the thirty-second annual ACM symposium on Theory of
  computing}}}\ (\bibinfo  {publisher} {ACM},\ \bibinfo {year}
  {2000})\BibitemShut {NoStop}%
\bibitem [{\citenamefont {Bru{\ss}}\ \emph {et~al.}(1998)\citenamefont
  {Bru{\ss}}, \citenamefont {DiVincenzo}, \citenamefont {Ekert}, \citenamefont
  {Fuchs}, \citenamefont {Macchiavello},\ and\ \citenamefont
  {Smolin}}]{bruss1998optimal}%
  \BibitemOpen
  \bibfield  {author} {\bibinfo {author} {\bibfnamefont {D.}~\bibnamefont
  {Bru{\ss}}}, \bibinfo {author} {\bibfnamefont {D.~P.}\ \bibnamefont
  {DiVincenzo}}, \bibinfo {author} {\bibfnamefont {A.}~\bibnamefont {Ekert}},
  \bibinfo {author} {\bibfnamefont {C.~A.}\ \bibnamefont {Fuchs}}, \bibinfo
  {author} {\bibfnamefont {C.}~\bibnamefont {Macchiavello}},\ and\ \bibinfo
  {author} {\bibfnamefont {J.~A.}\ \bibnamefont {Smolin}},\ }\bibfield  {title}
  {\bibinfo {title} {Optimal universal and state-dependent quantum cloning},\
  }\href {https://doi.org/10.1103/physreva.57.2368} {\bibfield  {journal}
  {\bibinfo  {journal} {Physical Review A}\ }\textbf {\bibinfo {volume} {57}},\
  \bibinfo {pages} {2368–2378} (\bibinfo {year} {1998})}\BibitemShut
  {NoStop}%
\bibitem [{\citenamefont {Bužek}\ and\ \citenamefont
  {Hillery}(1996)}]{buvzek1996quantum}%
  \BibitemOpen
  \bibfield  {author} {\bibinfo {author} {\bibfnamefont {V.}~\bibnamefont
  {Bužek}}\ and\ \bibinfo {author} {\bibfnamefont {M.}~\bibnamefont
  {Hillery}},\ }\bibfield  {title} {\bibinfo {title} {Quantum copying: Beyond
  the no-cloning theorem},\ }\href {https://doi.org/10.1103/physreva.54.1844}
  {\bibfield  {journal} {\bibinfo  {journal} {Physical Review A}\ }\textbf
  {\bibinfo {volume} {54}},\ \bibinfo {pages} {1844–1852} (\bibinfo {year}
  {1996})}\BibitemShut {NoStop}%
\bibitem [{\citenamefont {Lamas-Linares}\ \emph {et~al.}(2002)\citenamefont
  {Lamas-Linares}, \citenamefont {Simon}, \citenamefont {Howell},\ and\
  \citenamefont {Bouwmeester}}]{lamas2002experimental}%
  \BibitemOpen
  \bibfield  {author} {\bibinfo {author} {\bibfnamefont {A.}~\bibnamefont
  {Lamas-Linares}}, \bibinfo {author} {\bibfnamefont {C.}~\bibnamefont
  {Simon}}, \bibinfo {author} {\bibfnamefont {J.~C.}\ \bibnamefont {Howell}},\
  and\ \bibinfo {author} {\bibfnamefont {D.}~\bibnamefont {Bouwmeester}},\
  }\bibfield  {title} {\bibinfo {title} {Experimental quantum cloning of single
  photons},\ }\href {https://doi.org/10.1126/science.1068972} {\bibfield
  {journal} {\bibinfo  {journal} {Science}\ }\textbf {\bibinfo {volume}
  {296}},\ \bibinfo {pages} {712–714} (\bibinfo {year} {2002})}\BibitemShut
  {NoStop}%
\bibitem [{\citenamefont {Ricci}\ \emph {et~al.}(2004)\citenamefont {Ricci},
  \citenamefont {Sciarrino}, \citenamefont {Sias},\ and\ \citenamefont
  {De~Martini}}]{Ricci2004}%
  \BibitemOpen
  \bibfield  {author} {\bibinfo {author} {\bibfnamefont {M.}~\bibnamefont
  {Ricci}}, \bibinfo {author} {\bibfnamefont {F.}~\bibnamefont {Sciarrino}},
  \bibinfo {author} {\bibfnamefont {C.}~\bibnamefont {Sias}},\ and\ \bibinfo
  {author} {\bibfnamefont {F.}~\bibnamefont {De~Martini}},\ }\bibfield  {title}
  {\bibinfo {title} {Teleportation scheme implementing the universal optimal
  quantum cloning machine and the universal not gate},\ }\href
  {https://doi.org/10.1103/physrevlett.92.047901} {\bibfield  {journal}
  {\bibinfo  {journal} {Physical Review Letters}\ }\textbf {\bibinfo {volume}
  {92}},\ \bibinfo {pages} {047901} (\bibinfo {year} {2004})}\BibitemShut
  {NoStop}%
\bibitem [{\citenamefont {Scarani}\ \emph {et~al.}(2005)\citenamefont
  {Scarani}, \citenamefont {Iblisdir}, \citenamefont {Gisin},\ and\
  \citenamefont {Acín}}]{scarani2005quantum}%
  \BibitemOpen
  \bibfield  {author} {\bibinfo {author} {\bibfnamefont {V.}~\bibnamefont
  {Scarani}}, \bibinfo {author} {\bibfnamefont {S.}~\bibnamefont {Iblisdir}},
  \bibinfo {author} {\bibfnamefont {N.}~\bibnamefont {Gisin}},\ and\ \bibinfo
  {author} {\bibfnamefont {A.}~\bibnamefont {Acín}},\ }\bibfield  {title}
  {\bibinfo {title} {Quantum cloning},\ }\href
  {https://doi.org/10.1103/revmodphys.77.1225} {\bibfield  {journal} {\bibinfo
  {journal} {Reviews of Modern Physics}\ }\textbf {\bibinfo {volume} {77}},\
  \bibinfo {pages} {1225–1256} (\bibinfo {year} {2005})}\BibitemShut
  {NoStop}%
\bibitem [{\citenamefont {Durt}\ \emph {et~al.}(2005)\citenamefont {Durt},
  \citenamefont {Fiurášek},\ and\ \citenamefont {Cerf}}]{durt2005economical}%
  \BibitemOpen
  \bibfield  {author} {\bibinfo {author} {\bibfnamefont {T.}~\bibnamefont
  {Durt}}, \bibinfo {author} {\bibfnamefont {J.}~\bibnamefont {Fiurášek}},\
  and\ \bibinfo {author} {\bibfnamefont {N.~J.}\ \bibnamefont {Cerf}},\
  }\bibfield  {title} {\bibinfo {title} {Economical quantum cloning in any
  dimension},\ }\href {https://doi.org/10.1103/physreva.72.052322} {\bibfield
  {journal} {\bibinfo  {journal} {Physical Review A}\ }\textbf {\bibinfo
  {volume} {72}},\ \bibinfo {pages} {052322} (\bibinfo {year}
  {2005})}\BibitemShut {NoStop}%
\bibitem [{\citenamefont {Stanev}\ \emph {et~al.}(2023)\citenamefont {Stanev},
  \citenamefont {Spagnolo},\ and\ \citenamefont
  {Sciarrino}}]{stanev2023deterministic}%
  \BibitemOpen
  \bibfield  {author} {\bibinfo {author} {\bibfnamefont {D.}~\bibnamefont
  {Stanev}}, \bibinfo {author} {\bibfnamefont {N.}~\bibnamefont {Spagnolo}},\
  and\ \bibinfo {author} {\bibfnamefont {F.}~\bibnamefont {Sciarrino}},\
  }\bibfield  {title} {\bibinfo {title} {Deterministic optimal quantum cloning
  via a quantum-optical neural network},\ }\href
  {https://doi.org/10.1103/physrevresearch.5.013139} {\bibfield  {journal}
  {\bibinfo  {journal} {Physical Review Research}\ }\textbf {\bibinfo {volume}
  {5}},\ \bibinfo {pages} {013139} (\bibinfo {year} {2023})}\BibitemShut
  {NoStop}%
\bibitem [{\citenamefont {Zhang}\ \emph {et~al.}(2012)\citenamefont {Zhang},
  \citenamefont {Yu}, \citenamefont {Cao},\ and\ \citenamefont
  {Ye}}]{zhang2012optimal}%
  \BibitemOpen
  \bibfield  {author} {\bibinfo {author} {\bibfnamefont {W.-H.}\ \bibnamefont
  {Zhang}}, \bibinfo {author} {\bibfnamefont {L.-B.}\ \bibnamefont {Yu}},
  \bibinfo {author} {\bibfnamefont {Z.-L.}\ \bibnamefont {Cao}},\ and\ \bibinfo
  {author} {\bibfnamefont {L.}~\bibnamefont {Ye}},\ }\bibfield  {title}
  {\bibinfo {title} {Optimal cloning of two known nonorthogonal quantum
  states},\ }\href {https://doi.org/10.1103/physreva.86.022322} {\bibfield
  {journal} {\bibinfo  {journal} {Physical Review A}\ }\textbf {\bibinfo
  {volume} {86}},\ \bibinfo {pages} {022322} (\bibinfo {year}
  {2012})}\BibitemShut {NoStop}%
\bibitem [{\citenamefont {Coyle}\ \emph {et~al.}(2022)\citenamefont {Coyle},
  \citenamefont {Doosti}, \citenamefont {Kashefi},\ and\ \citenamefont
  {Kumar}}]{coyle2022progress}%
  \BibitemOpen
  \bibfield  {author} {\bibinfo {author} {\bibfnamefont {B.}~\bibnamefont
  {Coyle}}, \bibinfo {author} {\bibfnamefont {M.}~\bibnamefont {Doosti}},
  \bibinfo {author} {\bibfnamefont {E.}~\bibnamefont {Kashefi}},\ and\ \bibinfo
  {author} {\bibfnamefont {N.}~\bibnamefont {Kumar}},\ }\bibfield  {title}
  {\bibinfo {title} {Progress toward practical quantum cryptanalysis by
  variational quantum cloning},\ }\href
  {https://doi.org/10.1103/physreva.105.042604} {\bibfield  {journal} {\bibinfo
   {journal} {Physical Review A}\ }\textbf {\bibinfo {volume} {105}},\ \bibinfo
  {pages} {042604} (\bibinfo {year} {2022})}\BibitemShut {NoStop}%
\bibitem [{\citenamefont {Fan}\ \emph {et~al.}(2014)\citenamefont {Fan},
  \citenamefont {Wang}, \citenamefont {Jing}, \citenamefont {Yue},
  \citenamefont {Shi}, \citenamefont {Zhang},\ and\ \citenamefont
  {Mu}}]{fan2014quantum}%
  \BibitemOpen
  \bibfield  {author} {\bibinfo {author} {\bibfnamefont {H.}~\bibnamefont
  {Fan}}, \bibinfo {author} {\bibfnamefont {Y.-N.}\ \bibnamefont {Wang}},
  \bibinfo {author} {\bibfnamefont {L.}~\bibnamefont {Jing}}, \bibinfo {author}
  {\bibfnamefont {J.-D.}\ \bibnamefont {Yue}}, \bibinfo {author} {\bibfnamefont
  {H.-D.}\ \bibnamefont {Shi}}, \bibinfo {author} {\bibfnamefont {Y.-L.}\
  \bibnamefont {Zhang}},\ and\ \bibinfo {author} {\bibfnamefont {L.-Z.}\
  \bibnamefont {Mu}},\ }\bibfield  {title} {\bibinfo {title} {Quantum cloning
  machines and the applications},\ }\href
  {https://doi.org/10.1016/j.physrep.2014.06.004} {\bibfield  {journal}
  {\bibinfo  {journal} {Physics Reports}\ }\textbf {\bibinfo {volume} {544}},\
  \bibinfo {pages} {241–322} (\bibinfo {year} {2014})}\BibitemShut {NoStop}%
\bibitem [{\citenamefont {Wang}\ \emph {et~al.}(2011)\citenamefont {Wang},
  \citenamefont {Shi}, \citenamefont {Xiong}, \citenamefont {Jing},
  \citenamefont {Ren}, \citenamefont {Mu},\ and\ \citenamefont
  {Fan}}]{wang2011unified}%
  \BibitemOpen
  \bibfield  {author} {\bibinfo {author} {\bibfnamefont {Y.-N.}\ \bibnamefont
  {Wang}}, \bibinfo {author} {\bibfnamefont {H.-D.}\ \bibnamefont {Shi}},
  \bibinfo {author} {\bibfnamefont {Z.-X.}\ \bibnamefont {Xiong}}, \bibinfo
  {author} {\bibfnamefont {L.}~\bibnamefont {Jing}}, \bibinfo {author}
  {\bibfnamefont {X.-J.}\ \bibnamefont {Ren}}, \bibinfo {author} {\bibfnamefont
  {L.-Z.}\ \bibnamefont {Mu}},\ and\ \bibinfo {author} {\bibfnamefont
  {H.}~\bibnamefont {Fan}},\ }\bibfield  {title} {\bibinfo {title} {Unified
  universal quantum cloning machine and fidelities},\ }\href
  {https://doi.org/10.1103/physreva.84.034302} {\bibfield  {journal} {\bibinfo
  {journal} {Physical Review A}\ }\textbf {\bibinfo {volume} {84}},\ \bibinfo
  {pages} {034302} (\bibinfo {year} {2011})}\BibitemShut {NoStop}%
\bibitem [{\citenamefont {Xu}\ \emph {et~al.}(2008)\citenamefont {Xu},
  \citenamefont {Li}, \citenamefont {Chen}, \citenamefont {Zou},\ and\
  \citenamefont {Guo}}]{xu2008experimental}%
  \BibitemOpen
  \bibfield  {author} {\bibinfo {author} {\bibfnamefont {J.-S.}\ \bibnamefont
  {Xu}}, \bibinfo {author} {\bibfnamefont {C.-F.}\ \bibnamefont {Li}}, \bibinfo
  {author} {\bibfnamefont {L.}~\bibnamefont {Chen}}, \bibinfo {author}
  {\bibfnamefont {X.-B.}\ \bibnamefont {Zou}},\ and\ \bibinfo {author}
  {\bibfnamefont {G.-C.}\ \bibnamefont {Guo}},\ }\bibfield  {title} {\bibinfo
  {title} {Experimental realization of the optimal universal and
  phase-covariant quantum cloning machines},\ }\href
  {https://doi.org/10.1103/physreva.78.032322} {\bibfield  {journal} {\bibinfo
  {journal} {Physical Review A}\ }\textbf {\bibinfo {volume} {78}},\ \bibinfo
  {pages} {032322} (\bibinfo {year} {2008})}\BibitemShut {NoStop}%
\bibitem [{\citenamefont {Liu}\ \emph {et~al.}(2021)\citenamefont {Liu},
  \citenamefont {Lou}, \citenamefont {Chen},\ and\ \citenamefont
  {Jing}}]{liu2021all}%
  \BibitemOpen
  \bibfield  {author} {\bibinfo {author} {\bibfnamefont {S.}~\bibnamefont
  {Liu}}, \bibinfo {author} {\bibfnamefont {Y.}~\bibnamefont {Lou}}, \bibinfo
  {author} {\bibfnamefont {Y.}~\bibnamefont {Chen}},\ and\ \bibinfo {author}
  {\bibfnamefont {J.}~\bibnamefont {Jing}},\ }\href
  {https://doi.org/10.1103/physrevlett.126.060503} {\bibfield  {journal}
  {\bibinfo  {journal} {Physical Review Letters}\ }\textbf {\bibinfo {volume}
  {126}},\ \bibinfo {pages} {060503} (\bibinfo {year} {2021})}\BibitemShut
  {NoStop}%
\bibitem [{\citenamefont {Nagali}\ \emph {et~al.}(2009)\citenamefont {Nagali},
  \citenamefont {Sansoni}, \citenamefont {Sciarrino}, \citenamefont
  {De~Martini}, \citenamefont {Marrucci}, \citenamefont {Piccirillo},
  \citenamefont {Karimi},\ and\ \citenamefont {Santamato}}]{nagali2009optimal}%
  \BibitemOpen
  \bibfield  {author} {\bibinfo {author} {\bibfnamefont {E.}~\bibnamefont
  {Nagali}}, \bibinfo {author} {\bibfnamefont {L.}~\bibnamefont {Sansoni}},
  \bibinfo {author} {\bibfnamefont {F.}~\bibnamefont {Sciarrino}}, \bibinfo
  {author} {\bibfnamefont {F.}~\bibnamefont {De~Martini}}, \bibinfo {author}
  {\bibfnamefont {L.}~\bibnamefont {Marrucci}}, \bibinfo {author}
  {\bibfnamefont {B.}~\bibnamefont {Piccirillo}}, \bibinfo {author}
  {\bibfnamefont {E.}~\bibnamefont {Karimi}},\ and\ \bibinfo {author}
  {\bibfnamefont {E.}~\bibnamefont {Santamato}},\ }\bibfield  {title} {\bibinfo
  {title} {Optimal quantum cloning of orbital angular momentum photon qubits
  through hong–ou–mandel coalescence},\ }\href
  {https://doi.org/10.1038/nphoton.2009.214} {\bibfield  {journal} {\bibinfo
  {journal} {Nature Photonics}\ }\textbf {\bibinfo {volume} {3}},\ \bibinfo
  {pages} {720–723} (\bibinfo {year} {2009})}\BibitemShut {NoStop}%
\bibitem [{\citenamefont {Nagali}\ \emph {et~al.}(2010)\citenamefont {Nagali},
  \citenamefont {Giovannini}, \citenamefont {Marrucci}, \citenamefont
  {Slussarenko}, \citenamefont {Santamato},\ and\ \citenamefont
  {Sciarrino}}]{nagali2010experimental}%
  \BibitemOpen
  \bibfield  {author} {\bibinfo {author} {\bibfnamefont {E.}~\bibnamefont
  {Nagali}}, \bibinfo {author} {\bibfnamefont {D.}~\bibnamefont {Giovannini}},
  \bibinfo {author} {\bibfnamefont {L.}~\bibnamefont {Marrucci}}, \bibinfo
  {author} {\bibfnamefont {S.}~\bibnamefont {Slussarenko}}, \bibinfo {author}
  {\bibfnamefont {E.}~\bibnamefont {Santamato}},\ and\ \bibinfo {author}
  {\bibfnamefont {F.}~\bibnamefont {Sciarrino}},\ }\bibfield  {title} {\bibinfo
  {title} {Experimental optimal cloning of four-dimensional quantum states of
  photons},\ }\href {https://doi.org/10.1103/physrevlett.105.073602} {\bibfield
   {journal} {\bibinfo  {journal} {Physical Review Letters}\ }\textbf {\bibinfo
  {volume} {105}},\ \bibinfo {pages} {073602} (\bibinfo {year}
  {2010})}\BibitemShut {NoStop}%
\bibitem [{\citenamefont {Niu}\ and\ \citenamefont
  {Griffiths}(1999)}]{niu1999two}%
  \BibitemOpen
  \bibfield  {author} {\bibinfo {author} {\bibfnamefont {C.-S.}\ \bibnamefont
  {Niu}}\ and\ \bibinfo {author} {\bibfnamefont {R.~B.}\ \bibnamefont
  {Griffiths}},\ }\bibfield  {title} {\bibinfo {title} {Two-qubit copying
  machine for economical quantum eavesdropping},\ }\href
  {https://doi.org/10.1103/physreva.60.2764} {\bibfield  {journal} {\bibinfo
  {journal} {Physical Review A}\ }\textbf {\bibinfo {volume} {60}},\ \bibinfo
  {pages} {2764–2776} (\bibinfo {year} {1999})}\BibitemShut {NoStop}%
\bibitem [{\citenamefont {Duan}\ and\ \citenamefont {Guo}(1997)}]{duan1997two}%
  \BibitemOpen
  \bibfield  {author} {\bibinfo {author} {\bibfnamefont {L.-M.}\ \bibnamefont
  {Duan}}\ and\ \bibinfo {author} {\bibfnamefont {G.-C.}\ \bibnamefont {Guo}},\
  }\href@noop {} {\bibinfo {title} {Two non-orthogonal states can be cloned by
  a unitary-reduction process}} (\bibinfo {year} {1997}),\ \Eprint
  {https://arxiv.org/abs/quant-ph/9704020} {arXiv:quant-ph/9704020}
  \BibitemShut {NoStop}%
\bibitem [{\citenamefont {Preskill}(2018)}]{preskill2018quantum}%
  \BibitemOpen
  \bibfield  {author} {\bibinfo {author} {\bibfnamefont {J.}~\bibnamefont
  {Preskill}},\ }\bibfield  {title} {\bibinfo {title} {Quantum computing in the
  nisq era and beyond},\ }\href {https://doi.org/10.22331/q-2018-08-06-79}
  {\bibfield  {journal} {\bibinfo  {journal} {Quantum}\ }\textbf {\bibinfo
  {volume} {2}},\ \bibinfo {pages} {79} (\bibinfo {year} {2018})}\BibitemShut
  {NoStop}%
\bibitem [{\citenamefont {Biamonte}(2021)}]{biamonte2021universal}%
  \BibitemOpen
  \bibfield  {author} {\bibinfo {author} {\bibfnamefont {J.}~\bibnamefont
  {Biamonte}},\ }\bibfield  {title} {\bibinfo {title} {Universal variational
  quantum computation},\ }\href {https://doi.org/10.1103/physreva.103.l030401}
  {\bibfield  {journal} {\bibinfo  {journal} {Physical Review A}\ }\textbf
  {\bibinfo {volume} {103}},\ \bibinfo {pages} {L030401} (\bibinfo {year}
  {2021})}\BibitemShut {NoStop}%
\bibitem [{\citenamefont {Cerezo}\ \emph {et~al.}(2021)\citenamefont {Cerezo},
  \citenamefont {Arrasmith}, \citenamefont {Babbush}, \citenamefont {Benjamin},
  \citenamefont {Endo}, \citenamefont {Fujii}, \citenamefont {McClean},
  \citenamefont {Mitarai}, \citenamefont {Yuan}, \citenamefont {Cincio},\ and\
  \citenamefont {Coles}}]{cerezo2021variational}%
  \BibitemOpen
  \bibfield  {author} {\bibinfo {author} {\bibfnamefont {M.}~\bibnamefont
  {Cerezo}}, \bibinfo {author} {\bibfnamefont {A.}~\bibnamefont {Arrasmith}},
  \bibinfo {author} {\bibfnamefont {R.}~\bibnamefont {Babbush}}, \bibinfo
  {author} {\bibfnamefont {S.~C.}\ \bibnamefont {Benjamin}}, \bibinfo {author}
  {\bibfnamefont {S.}~\bibnamefont {Endo}}, \bibinfo {author} {\bibfnamefont
  {K.}~\bibnamefont {Fujii}}, \bibinfo {author} {\bibfnamefont {J.~R.}\
  \bibnamefont {McClean}}, \bibinfo {author} {\bibfnamefont {K.}~\bibnamefont
  {Mitarai}}, \bibinfo {author} {\bibfnamefont {X.}~\bibnamefont {Yuan}},
  \bibinfo {author} {\bibfnamefont {L.}~\bibnamefont {Cincio}},\ and\ \bibinfo
  {author} {\bibfnamefont {P.~J.}\ \bibnamefont {Coles}},\ }\bibfield  {title}
  {\bibinfo {title} {Variational quantum algorithms},\ }\href
  {https://doi.org/10.1038/s42254-021-00348-9} {\bibfield  {journal} {\bibinfo
  {journal} {Nature Reviews Physics}\ }\textbf {\bibinfo {volume} {3}},\
  \bibinfo {pages} {625–644} (\bibinfo {year} {2021})}\BibitemShut {NoStop}%
\bibitem [{\citenamefont {Wecker}\ \emph {et~al.}(2015)\citenamefont {Wecker},
  \citenamefont {Hastings},\ and\ \citenamefont {Troyer}}]{wecker2015progress}%
  \BibitemOpen
  \bibfield  {author} {\bibinfo {author} {\bibfnamefont {D.}~\bibnamefont
  {Wecker}}, \bibinfo {author} {\bibfnamefont {M.~B.}\ \bibnamefont
  {Hastings}},\ and\ \bibinfo {author} {\bibfnamefont {M.}~\bibnamefont
  {Troyer}},\ }\bibfield  {title} {\bibinfo {title} {Progress towards practical
  quantum variational algorithms},\ }\href
  {https://doi.org/10.1103/physreva.92.042303} {\bibfield  {journal} {\bibinfo
  {journal} {Physical Review A}\ }\textbf {\bibinfo {volume} {92}},\ \bibinfo
  {pages} {042303} (\bibinfo {year} {2015})}\BibitemShut {NoStop}%
\bibitem [{\citenamefont {Endo}\ \emph {et~al.}(2021)\citenamefont {Endo},
  \citenamefont {Cai}, \citenamefont {Benjamin},\ and\ \citenamefont
  {Yuan}}]{endo2021hybrid}%
  \BibitemOpen
  \bibfield  {author} {\bibinfo {author} {\bibfnamefont {S.}~\bibnamefont
  {Endo}}, \bibinfo {author} {\bibfnamefont {Z.}~\bibnamefont {Cai}}, \bibinfo
  {author} {\bibfnamefont {S.~C.}\ \bibnamefont {Benjamin}},\ and\ \bibinfo
  {author} {\bibfnamefont {X.}~\bibnamefont {Yuan}},\ }\bibfield  {title}
  {\bibinfo {title} {Hybrid quantum-classical algorithms and quantum error
  mitigation},\ }\href {https://doi.org/10.7566/jpsj.90.032001} {\bibfield
  {journal} {\bibinfo  {journal} {Journal of the Physical Society of Japan}\
  }\textbf {\bibinfo {volume} {90}},\ \bibinfo {pages} {032001} (\bibinfo
  {year} {2021})}\BibitemShut {NoStop}%
\bibitem [{\citenamefont {McClean}\ \emph {et~al.}(2016)\citenamefont
  {McClean}, \citenamefont {Romero}, \citenamefont {Babbush},\ and\
  \citenamefont {Aspuru-Guzik}}]{mcclean2016theory}%
  \BibitemOpen
  \bibfield  {author} {\bibinfo {author} {\bibfnamefont {J.~R.}\ \bibnamefont
  {McClean}}, \bibinfo {author} {\bibfnamefont {J.}~\bibnamefont {Romero}},
  \bibinfo {author} {\bibfnamefont {R.}~\bibnamefont {Babbush}},\ and\ \bibinfo
  {author} {\bibfnamefont {A.}~\bibnamefont {Aspuru-Guzik}},\ }\bibfield
  {title} {\bibinfo {title} {The theory of variational hybrid quantum-classical
  algorithms},\ }\href {https://doi.org/10.1088/1367-2630/18/2/023023}
  {\bibfield  {journal} {\bibinfo  {journal} {New Journal of Physics}\ }\textbf
  {\bibinfo {volume} {18}},\ \bibinfo {pages} {023023} (\bibinfo {year}
  {2016})}\BibitemShut {NoStop}%
\bibitem [{\citenamefont {Peruzzo}\ \emph {et~al.}(2014)\citenamefont
  {Peruzzo}, \citenamefont {McClean}, \citenamefont {Shadbolt}, \citenamefont
  {Yung}, \citenamefont {Zhou}, \citenamefont {Love}, \citenamefont
  {Aspuru-Guzik},\ and\ \citenamefont {O’Brien}}]{peruzzo2014variational}%
  \BibitemOpen
  \bibfield  {author} {\bibinfo {author} {\bibfnamefont {A.}~\bibnamefont
  {Peruzzo}}, \bibinfo {author} {\bibfnamefont {J.}~\bibnamefont {McClean}},
  \bibinfo {author} {\bibfnamefont {P.}~\bibnamefont {Shadbolt}}, \bibinfo
  {author} {\bibfnamefont {M.-H.}\ \bibnamefont {Yung}}, \bibinfo {author}
  {\bibfnamefont {X.-Q.}\ \bibnamefont {Zhou}}, \bibinfo {author}
  {\bibfnamefont {P.~J.}\ \bibnamefont {Love}}, \bibinfo {author}
  {\bibfnamefont {A.}~\bibnamefont {Aspuru-Guzik}},\ and\ \bibinfo {author}
  {\bibfnamefont {J.~L.}\ \bibnamefont {O’Brien}},\ }\bibfield  {title}
  {\bibinfo {title} {A variational eigenvalue solver on a photonic quantum
  processor},\ }\href {https://doi.org/10.1038/ncomms5213} {\bibfield
  {journal} {\bibinfo  {journal} {Nature Communications}\ }\textbf {\bibinfo
  {volume} {5}},\ \bibinfo {pages} {4213} (\bibinfo {year} {2014})}\BibitemShut
  {NoStop}%
\bibitem [{\citenamefont {Morales}\ \emph {et~al.}(2018)\citenamefont
  {Morales}, \citenamefont {Tlyachev},\ and\ \citenamefont
  {Biamonte}}]{morales2018variational}%
  \BibitemOpen
  \bibfield  {author} {\bibinfo {author} {\bibfnamefont {M.~E.~S.}\
  \bibnamefont {Morales}}, \bibinfo {author} {\bibfnamefont {T.}~\bibnamefont
  {Tlyachev}},\ and\ \bibinfo {author} {\bibfnamefont {J.}~\bibnamefont
  {Biamonte}},\ }\bibfield  {title} {\bibinfo {title} {Variational learning of
  grover’s quantum search algorithm},\ }\href
  {https://doi.org/10.1103/physreva.98.062333} {\bibfield  {journal} {\bibinfo
  {journal} {Physical Review A}\ }\textbf {\bibinfo {volume} {98}},\ \bibinfo
  {pages} {062333} (\bibinfo {year} {2018})}\BibitemShut {NoStop}%
\bibitem [{\citenamefont {Khatri}\ \emph {et~al.}(2019)\citenamefont {Khatri},
  \citenamefont {LaRose}, \citenamefont {Poremba}, \citenamefont {Cincio},
  \citenamefont {Sornborger},\ and\ \citenamefont {Coles}}]{khatri2019quantum}%
  \BibitemOpen
  \bibfield  {author} {\bibinfo {author} {\bibfnamefont {S.}~\bibnamefont
  {Khatri}}, \bibinfo {author} {\bibfnamefont {R.}~\bibnamefont {LaRose}},
  \bibinfo {author} {\bibfnamefont {A.}~\bibnamefont {Poremba}}, \bibinfo
  {author} {\bibfnamefont {L.}~\bibnamefont {Cincio}}, \bibinfo {author}
  {\bibfnamefont {A.~T.}\ \bibnamefont {Sornborger}},\ and\ \bibinfo {author}
  {\bibfnamefont {P.~J.}\ \bibnamefont {Coles}},\ }\bibfield  {title} {\bibinfo
  {title} {Quantum-assisted quantum compiling},\ }\href
  {https://doi.org/10.22331/q-2019-05-13-140} {\bibfield  {journal} {\bibinfo
  {journal} {Quantum}\ }\textbf {\bibinfo {volume} {3}},\ \bibinfo {pages}
  {140} (\bibinfo {year} {2019})}\BibitemShut {NoStop}%
\bibitem [{\citenamefont {Jones}\ and\ \citenamefont
  {Benjamin}(2022)}]{jones2022robust}%
  \BibitemOpen
  \bibfield  {author} {\bibinfo {author} {\bibfnamefont {T.}~\bibnamefont
  {Jones}}\ and\ \bibinfo {author} {\bibfnamefont {S.~C.}\ \bibnamefont
  {Benjamin}},\ }\bibfield  {title} {\bibinfo {title} {Robust quantum
  compilation and circuit optimisation via energy minimisation},\ }\href
  {https://doi.org/10.22331/q-2022-01-24-628} {\bibfield  {journal} {\bibinfo
  {journal} {Quantum}\ }\textbf {\bibinfo {volume} {6}},\ \bibinfo {pages}
  {628} (\bibinfo {year} {2022})}\BibitemShut {NoStop}%
\bibitem [{\citenamefont {Bravo-Prieto}\ \emph {et~al.}(2023)\citenamefont
  {Bravo-Prieto}, \citenamefont {LaRose}, \citenamefont {Cerezo}, \citenamefont
  {Subasi}, \citenamefont {Cincio},\ and\ \citenamefont
  {Coles}}]{bravo2023variational}%
  \BibitemOpen
  \bibfield  {author} {\bibinfo {author} {\bibfnamefont {C.}~\bibnamefont
  {Bravo-Prieto}}, \bibinfo {author} {\bibfnamefont {R.}~\bibnamefont
  {LaRose}}, \bibinfo {author} {\bibfnamefont {M.}~\bibnamefont {Cerezo}},
  \bibinfo {author} {\bibfnamefont {Y.}~\bibnamefont {Subasi}}, \bibinfo
  {author} {\bibfnamefont {L.}~\bibnamefont {Cincio}},\ and\ \bibinfo {author}
  {\bibfnamefont {P.~J.}\ \bibnamefont {Coles}},\ }\bibfield  {title} {\bibinfo
  {title} {Variational quantum linear solver},\ }\href
  {https://doi.org/10.22331/q-2023-11-22-1188} {\bibfield  {journal} {\bibinfo
  {journal} {Quantum}\ }\textbf {\bibinfo {volume} {7}},\ \bibinfo {pages}
  {1188} (\bibinfo {year} {2023})}\BibitemShut {NoStop}%
\bibitem [{\citenamefont {Arrasmith}\ \emph {et~al.}(2019)\citenamefont
  {Arrasmith}, \citenamefont {Cincio}, \citenamefont {Sornborger},
  \citenamefont {Zurek},\ and\ \citenamefont
  {Coles}}]{arrasmith2019variational}%
  \BibitemOpen
  \bibfield  {author} {\bibinfo {author} {\bibfnamefont {A.}~\bibnamefont
  {Arrasmith}}, \bibinfo {author} {\bibfnamefont {L.}~\bibnamefont {Cincio}},
  \bibinfo {author} {\bibfnamefont {A.~T.}\ \bibnamefont {Sornborger}},
  \bibinfo {author} {\bibfnamefont {W.~H.}\ \bibnamefont {Zurek}},\ and\
  \bibinfo {author} {\bibfnamefont {P.~J.}\ \bibnamefont {Coles}},\ }\bibfield
  {title} {\bibinfo {title} {Variational consistent histories as a hybrid
  algorithm for quantum foundations},\ }\href
  {https://doi.org/10.1038/s41467-019-11417-0} {\bibfield  {journal} {\bibinfo
  {journal} {Nature Communications}\ }\textbf {\bibinfo {volume} {10}},\
  \bibinfo {pages} {3438} (\bibinfo {year} {2019})}\BibitemShut {NoStop}%
\bibitem [{\citenamefont {Wittek}(2014)}]{wittek2014quantum}%
  \BibitemOpen
  \bibfield  {author} {\bibinfo {author} {\bibfnamefont {P.}~\bibnamefont
  {Wittek}},\ }\href
  {https://www.sciencedirect.com/book/9780128009536/quantum-machine-learning}
  {\emph {\bibinfo {title} {Quantum machine learning: what quantum computing
  means to data mining}}}\ (\bibinfo  {publisher} {Academic Press},\ \bibinfo
  {year} {2014})\BibitemShut {NoStop}%
\bibitem [{\citenamefont {Biamonte}\ \emph {et~al.}(2017)\citenamefont
  {Biamonte}, \citenamefont {Wittek}, \citenamefont {Pancotti}, \citenamefont
  {Rebentrost}, \citenamefont {Wiebe},\ and\ \citenamefont
  {Lloyd}}]{biamonte2017quantum}%
  \BibitemOpen
  \bibfield  {author} {\bibinfo {author} {\bibfnamefont {J.}~\bibnamefont
  {Biamonte}}, \bibinfo {author} {\bibfnamefont {P.}~\bibnamefont {Wittek}},
  \bibinfo {author} {\bibfnamefont {N.}~\bibnamefont {Pancotti}}, \bibinfo
  {author} {\bibfnamefont {P.}~\bibnamefont {Rebentrost}}, \bibinfo {author}
  {\bibfnamefont {N.}~\bibnamefont {Wiebe}},\ and\ \bibinfo {author}
  {\bibfnamefont {S.}~\bibnamefont {Lloyd}},\ }\bibfield  {title} {\bibinfo
  {title} {Quantum machine learning},\ }\href
  {https://doi.org/10.1038/nature23474} {\bibfield  {journal} {\bibinfo
  {journal} {Nature}\ }\textbf {\bibinfo {volume} {549}},\ \bibinfo {pages}
  {195–202} (\bibinfo {year} {2017})}\BibitemShut {NoStop}%
\bibitem [{\citenamefont {Kopczyk}(2018)}]{kopczyk2018quantum}%
  \BibitemOpen
  \bibfield  {author} {\bibinfo {author} {\bibfnamefont {D.}~\bibnamefont
  {Kopczyk}},\ }\href@noop {} {\bibinfo {title} {Quantum machine learning for
  data scientists}} (\bibinfo {year} {2018}),\ \Eprint
  {https://arxiv.org/abs/1804.10068} {arXiv:1804.10068 [quant-ph]} \BibitemShut
  {NoStop}%
\bibitem [{\citenamefont {Schuld}\ and\ \citenamefont
  {Petruccione}(2018)}]{schuld2018supervised}%
  \BibitemOpen
  \bibfield  {author} {\bibinfo {author} {\bibfnamefont {M.}~\bibnamefont
  {Schuld}}\ and\ \bibinfo {author} {\bibfnamefont {F.}~\bibnamefont
  {Petruccione}},\ }\href
  {https://link.springer.com/book/10.1007/978-3-319-96424-9} {\emph {\bibinfo
  {title} {Supervised learning with quantum computers}}},\ Vol.~\bibinfo
  {volume} {17}\ (\bibinfo  {publisher} {Springer},\ \bibinfo {year}
  {2018})\BibitemShut {NoStop}%
\bibitem [{\citenamefont {Peters}\ \emph {et~al.}(2003)\citenamefont {Peters},
  \citenamefont {Altepeter}, \citenamefont {Jeffrey}, \citenamefont
  {Branning},\ and\ \citenamefont {Kwiat}}]{peters2003precise}%
  \BibitemOpen
  \bibfield  {author} {\bibinfo {author} {\bibfnamefont {N.}~\bibnamefont
  {Peters}}, \bibinfo {author} {\bibfnamefont {J.}~\bibnamefont {Altepeter}},
  \bibinfo {author} {\bibfnamefont {E.}~\bibnamefont {Jeffrey}}, \bibinfo
  {author} {\bibfnamefont {D.}~\bibnamefont {Branning}},\ and\ \bibinfo
  {author} {\bibfnamefont {P.}~\bibnamefont {Kwiat}},\ }\bibfield  {title}
  {\bibinfo {title} {Precise creation, characterization, and manipulation of
  single optical qubits},\ }\href {https://doi.org/10.26421/QIC3.s-4}
  {\bibfield  {journal} {\bibinfo  {journal} {Quantum Information and
  Computation}\ }\textbf {\bibinfo {volume} {3}},\ \bibinfo {pages} {503}
  (\bibinfo {year} {2003})}\BibitemShut {NoStop}%
\bibitem [{\citenamefont {Cacciapuoti}\ \emph {et~al.}(2019)\citenamefont
  {Cacciapuoti}, \citenamefont {Caleffi}, \citenamefont {Tafuri}, \citenamefont
  {Cataliotti}, \citenamefont {Gherardini},\ and\ \citenamefont
  {Bianchi}}]{cacciapuoti2019quantum}%
  \BibitemOpen
  \bibfield  {author} {\bibinfo {author} {\bibfnamefont {A.~S.}\ \bibnamefont
  {Cacciapuoti}}, \bibinfo {author} {\bibfnamefont {M.}~\bibnamefont
  {Caleffi}}, \bibinfo {author} {\bibfnamefont {F.}~\bibnamefont {Tafuri}},
  \bibinfo {author} {\bibfnamefont {F.~S.}\ \bibnamefont {Cataliotti}},
  \bibinfo {author} {\bibfnamefont {S.}~\bibnamefont {Gherardini}},\ and\
  \bibinfo {author} {\bibfnamefont {G.}~\bibnamefont {Bianchi}},\ }\bibfield
  {title} {\bibinfo {title} {Quantum internet: Networking challenges in
  distributed quantum computing},\ }\href
  {https://doi.org/10.1109/MNET.001.1900092} {\bibfield  {journal} {\bibinfo
  {journal} {IEEE Network}\ }\textbf {\bibinfo {volume} {34}},\ \bibinfo
  {pages} {137} (\bibinfo {year} {2019})}\BibitemShut {NoStop}%
\bibitem [{\citenamefont {Jašek}\ \emph {et~al.}(2019)\citenamefont {Jašek},
  \citenamefont {Jiráková}, \citenamefont {Bartkiewicz}, \citenamefont
  {Černoch}, \citenamefont {F\"{u}rst},\ and\ \citenamefont
  {Lemr}}]{javsek2019experimental}%
  \BibitemOpen
  \bibfield  {author} {\bibinfo {author} {\bibfnamefont {J.}~\bibnamefont
  {Jašek}}, \bibinfo {author} {\bibfnamefont {K.}~\bibnamefont {Jiráková}},
  \bibinfo {author} {\bibfnamefont {K.}~\bibnamefont {Bartkiewicz}}, \bibinfo
  {author} {\bibfnamefont {A.}~\bibnamefont {Černoch}}, \bibinfo {author}
  {\bibfnamefont {T.}~\bibnamefont {F\"{u}rst}},\ and\ \bibinfo {author}
  {\bibfnamefont {K.}~\bibnamefont {Lemr}},\ }\bibfield  {title} {\bibinfo
  {title} {Experimental hybrid quantum-classical reinforcement learning by
  boson sampling: how to train a quantum cloner},\ }\href
  {https://doi.org/10.1364/oe.27.032454} {\bibfield  {journal} {\bibinfo
  {journal} {Optics Express}\ }\textbf {\bibinfo {volume} {27}},\ \bibinfo
  {pages} {32454} (\bibinfo {year} {2019})}\BibitemShut {NoStop}%
\bibitem [{\citenamefont {Sciarrino}\ and\ \citenamefont
  {De~Martini}(2007)}]{sciarrino2007implementation}%
  \BibitemOpen
  \bibfield  {author} {\bibinfo {author} {\bibfnamefont {F.}~\bibnamefont
  {Sciarrino}}\ and\ \bibinfo {author} {\bibfnamefont {F.}~\bibnamefont
  {De~Martini}},\ }\bibfield  {title} {\bibinfo {title} {Implementation of
  optimal phase-covariant cloning machines},\ }\href
  {https://doi.org/10.1103/physreva.76.012330} {\bibfield  {journal} {\bibinfo
  {journal} {Physical Review A}\ }\textbf {\bibinfo {volume} {76}},\ \bibinfo
  {pages} {012330} (\bibinfo {year} {2007})}\BibitemShut {NoStop}%
\bibitem [{\citenamefont {Soubusta}\ \emph {et~al.}(2007)\citenamefont
  {Soubusta}, \citenamefont {Bart\ifmmode \mathring{u}\else
  \r{u}\fi{}\ifmmode~\check{s}\else \v{s}\fi{}kov\'a}, \citenamefont
  {\ifmmode~\check{C}\else \v{C}\fi{}ernoch}, \citenamefont
  {Fiur\'a\ifmmode~\check{s}\else \v{s}\fi{}ek},\ and\ \citenamefont
  {Du\ifmmode~\check{s}\else \v{s}\fi{}ek}}]{soubusta2007several}%
  \BibitemOpen
  \bibfield  {author} {\bibinfo {author} {\bibfnamefont {J.}~\bibnamefont
  {Soubusta}}, \bibinfo {author} {\bibfnamefont {L.}~\bibnamefont {Bart\ifmmode
  \mathring{u}\else \r{u}\fi{}\ifmmode~\check{s}\else \v{s}\fi{}kov\'a}},
  \bibinfo {author} {\bibfnamefont {A.}~\bibnamefont {\ifmmode~\check{C}\else
  \v{C}\fi{}ernoch}}, \bibinfo {author} {\bibfnamefont {J.}~\bibnamefont
  {Fiur\'a\ifmmode~\check{s}\else \v{s}\fi{}ek}},\ and\ \bibinfo {author}
  {\bibfnamefont {M.}~\bibnamefont {Du\ifmmode~\check{s}\else \v{s}\fi{}ek}},\
  }\bibfield  {title} {\bibinfo {title} {Several experimental realizations of
  symmetric phase-covariant quantum cloners of single-photon qubits},\ }\href
  {https://doi.org/10.1103/physreva.76.042318} {\bibfield  {journal} {\bibinfo
  {journal} {Physical Review A}\ }\textbf {\bibinfo {volume} {76}},\ \bibinfo
  {pages} {042318} (\bibinfo {year} {2007})}\BibitemShut {NoStop}%
\bibitem [{\citenamefont {Bart\ifmmode \mathring{u}\else
  \r{u}\fi{}\ifmmode~\check{s}\else \v{s}\fi{}kov\'a}\ \emph
  {et~al.}(2007)\citenamefont {Bart\ifmmode \mathring{u}\else
  \r{u}\fi{}\ifmmode~\check{s}\else \v{s}\fi{}kov\'a}, \citenamefont
  {Du\ifmmode~\check{s}\else \v{s}\fi{}ek}, \citenamefont
  {\ifmmode~\check{C}\else \v{C}\fi{}ernoch}, \citenamefont {Soubusta},\ and\
  \citenamefont {Fiur\'a\ifmmode~\check{s}\else
  \v{s}\fi{}ek}}]{bartuuvskova2007fiber}%
  \BibitemOpen
  \bibfield  {author} {\bibinfo {author} {\bibfnamefont {L.}~\bibnamefont
  {Bart\ifmmode \mathring{u}\else \r{u}\fi{}\ifmmode~\check{s}\else
  \v{s}\fi{}kov\'a}}, \bibinfo {author} {\bibfnamefont {M.}~\bibnamefont
  {Du\ifmmode~\check{s}\else \v{s}\fi{}ek}}, \bibinfo {author} {\bibfnamefont
  {A.}~\bibnamefont {\ifmmode~\check{C}\else \v{C}\fi{}ernoch}}, \bibinfo
  {author} {\bibfnamefont {J.}~\bibnamefont {Soubusta}},\ and\ \bibinfo
  {author} {\bibfnamefont {J.}~\bibnamefont {Fiur\'a\ifmmode~\check{s}\else
  \v{s}\fi{}ek}},\ }\bibfield  {title} {\bibinfo {title} {Fiber-optics
  implementation of an asymmetric phase-covariant quantum cloner},\ }\href
  {https://doi.org/10.1103/physrevlett.99.120505} {\bibfield  {journal}
  {\bibinfo  {journal} {Physical Review Letters}\ }\textbf {\bibinfo {volume}
  {99}},\ \bibinfo {pages} {120505} (\bibinfo {year} {2007})}\BibitemShut
  {NoStop}%
\bibitem [{\citenamefont {\ifmmode~\check{C}\else \v{C}\fi{}ernoch}\ \emph
  {et~al.}(2006)\citenamefont {\ifmmode~\check{C}\else \v{C}\fi{}ernoch},
  \citenamefont {Bart\ifmmode \mathring{u}\else
  \r{u}\fi{}\ifmmode~\check{s}\else \v{s}\fi{}kov\'a}, \citenamefont
  {Soubusta}, \citenamefont {Je\ifmmode~\check{z}\else \v{z}\fi{}ek},
  \citenamefont {Fiur\'a\ifmmode~\check{s}\else \v{s}\fi{}ek},\ and\
  \citenamefont {Du\ifmmode~\check{s}\else
  \v{s}\fi{}ek}}]{vcernoch2006experimental}%
  \BibitemOpen
  \bibfield  {author} {\bibinfo {author} {\bibfnamefont {A.}~\bibnamefont
  {\ifmmode~\check{C}\else \v{C}\fi{}ernoch}}, \bibinfo {author} {\bibfnamefont
  {L.}~\bibnamefont {Bart\ifmmode \mathring{u}\else
  \r{u}\fi{}\ifmmode~\check{s}\else \v{s}\fi{}kov\'a}}, \bibinfo {author}
  {\bibfnamefont {J.}~\bibnamefont {Soubusta}}, \bibinfo {author}
  {\bibfnamefont {M.}~\bibnamefont {Je\ifmmode~\check{z}\else \v{z}\fi{}ek}},
  \bibinfo {author} {\bibfnamefont {J.}~\bibnamefont
  {Fiur\'a\ifmmode~\check{s}\else \v{s}\fi{}ek}},\ and\ \bibinfo {author}
  {\bibfnamefont {M.}~\bibnamefont {Du\ifmmode~\check{s}\else \v{s}\fi{}ek}},\
  }\bibfield  {title} {\bibinfo {title} {Experimental phase-covariant cloning
  of polarization states of single photons},\ }\href
  {https://doi.org/10.1103/physreva.74.042327} {\bibfield  {journal} {\bibinfo
  {journal} {Physical Review A}\ }\textbf {\bibinfo {volume} {74}},\ \bibinfo
  {pages} {042327} (\bibinfo {year} {2006})}\BibitemShut {NoStop}%
\bibitem [{\citenamefont {D’Ariano}\ and\ \citenamefont
  {Macchiavello}(2003)}]{d2003optimal}%
  \BibitemOpen
  \bibfield  {author} {\bibinfo {author} {\bibfnamefont {G.~M.}\ \bibnamefont
  {D’Ariano}}\ and\ \bibinfo {author} {\bibfnamefont {C.}~\bibnamefont
  {Macchiavello}},\ }\bibfield  {title} {\bibinfo {title} {Optimal
  phase-covariant cloning for qubits and qutrits},\ }\href
  {https://doi.org/10.1103/physreva.67.042306} {\bibfield  {journal} {\bibinfo
  {journal} {Physical Review A}\ }\textbf {\bibinfo {volume} {67}},\ \bibinfo
  {pages} {042306} (\bibinfo {year} {2003})}\BibitemShut {NoStop}%
\bibitem [{\citenamefont {Chen}\ \emph {et~al.}(2007)\citenamefont {Chen},
  \citenamefont {Zhou}, \citenamefont {Suter},\ and\ \citenamefont
  {Du}}]{chen2007experimental}%
  \BibitemOpen
  \bibfield  {author} {\bibinfo {author} {\bibfnamefont {H.}~\bibnamefont
  {Chen}}, \bibinfo {author} {\bibfnamefont {X.}~\bibnamefont {Zhou}}, \bibinfo
  {author} {\bibfnamefont {D.}~\bibnamefont {Suter}},\ and\ \bibinfo {author}
  {\bibfnamefont {J.}~\bibnamefont {Du}},\ }\bibfield  {title} {\bibinfo
  {title} {Experimental realization of $1 \rightarrow 2$ asymmetric
  phase-covariant quantum cloning},\ }\href
  {https://doi.org/10.1103/physreva.75.012317} {\bibfield  {journal} {\bibinfo
  {journal} {Physical Review A}\ }\textbf {\bibinfo {volume} {75}},\ \bibinfo
  {pages} {012317} (\bibinfo {year} {2007})}\BibitemShut {NoStop}%
\bibitem [{\citenamefont {Fiurášek}(2003)}]{fiuravsek2003optical}%
  \BibitemOpen
  \bibfield  {author} {\bibinfo {author} {\bibfnamefont {J.}~\bibnamefont
  {Fiurášek}},\ }\bibfield  {title} {\bibinfo {title} {Optical
  implementations of the optimal phase-covariant quantum cloning machine},\
  }\href {https://doi.org/10.1103/physreva.67.052314} {\bibfield  {journal}
  {\bibinfo  {journal} {Physical Review A}\ }\textbf {\bibinfo {volume} {67}},\
  \bibinfo {pages} {052314} (\bibinfo {year} {2003})}\BibitemShut {NoStop}%
\bibitem [{\citenamefont {Sciarrino}\ and\ \citenamefont
  {De~Martini}(2005)}]{sciarrino2005realization}%
  \BibitemOpen
  \bibfield  {author} {\bibinfo {author} {\bibfnamefont {F.}~\bibnamefont
  {Sciarrino}}\ and\ \bibinfo {author} {\bibfnamefont {F.}~\bibnamefont
  {De~Martini}},\ }\bibfield  {title} {\bibinfo {title} {Realization of the
  optimal phase-covariant quantum cloning machine},\ }\href
  {https://doi.org/10.1103/physreva.72.062313} {\bibfield  {journal} {\bibinfo
  {journal} {Physical Review A}\ }\textbf {\bibinfo {volume} {72}},\ \bibinfo
  {pages} {062313} (\bibinfo {year} {2005})}\BibitemShut {NoStop}%
\bibitem [{\citenamefont {Bru{\ss}}\ \emph {et~al.}(2000)\citenamefont
  {Bru{\ss}}, \citenamefont {Cinchetti}, \citenamefont {Mauro~D’Ariano},\
  and\ \citenamefont {Macchiavello}}]{bruss2000phase}%
  \BibitemOpen
  \bibfield  {author} {\bibinfo {author} {\bibfnamefont {D.}~\bibnamefont
  {Bru{\ss}}}, \bibinfo {author} {\bibfnamefont {M.}~\bibnamefont {Cinchetti}},
  \bibinfo {author} {\bibfnamefont {G.}~\bibnamefont {Mauro~D’Ariano}},\ and\
  \bibinfo {author} {\bibfnamefont {C.}~\bibnamefont {Macchiavello}},\
  }\bibfield  {title} {\bibinfo {title} {Phase-covariant quantum cloning},\
  }\href {https://doi.org/10.1103/physreva.62.012302} {\bibfield  {journal}
  {\bibinfo  {journal} {Physical Review A}\ }\textbf {\bibinfo {volume} {62}},\
  \bibinfo {pages} {012302} (\bibinfo {year} {2000})}\BibitemShut {NoStop}%
\bibitem [{\citenamefont {Spagnolo}\ \emph {et~al.}(2022)\citenamefont
  {Spagnolo}, \citenamefont {Morris}, \citenamefont {Piacentini}, \citenamefont
  {Antesberger}, \citenamefont {Massa}, \citenamefont {Crespi}, \citenamefont
  {Ceccarelli}, \citenamefont {Osellame},\ and\ \citenamefont
  {Walther}}]{spagnolo2022experimental}%
  \BibitemOpen
  \bibfield  {author} {\bibinfo {author} {\bibfnamefont {M.}~\bibnamefont
  {Spagnolo}}, \bibinfo {author} {\bibfnamefont {J.}~\bibnamefont {Morris}},
  \bibinfo {author} {\bibfnamefont {S.}~\bibnamefont {Piacentini}}, \bibinfo
  {author} {\bibfnamefont {M.}~\bibnamefont {Antesberger}}, \bibinfo {author}
  {\bibfnamefont {F.}~\bibnamefont {Massa}}, \bibinfo {author} {\bibfnamefont
  {A.}~\bibnamefont {Crespi}}, \bibinfo {author} {\bibfnamefont
  {F.}~\bibnamefont {Ceccarelli}}, \bibinfo {author} {\bibfnamefont
  {R.}~\bibnamefont {Osellame}},\ and\ \bibinfo {author} {\bibfnamefont
  {P.}~\bibnamefont {Walther}},\ }\bibfield  {title} {\bibinfo {title}
  {Experimental photonic quantum memristor},\ }\href
  {https://doi.org/10.1038/s41566-022-00973-5} {\bibfield  {journal} {\bibinfo
  {journal} {Nature Photonics}\ }\textbf {\bibinfo {volume} {16}},\ \bibinfo
  {pages} {318–323} (\bibinfo {year} {2022})}\BibitemShut {NoStop}%
\bibitem [{\citenamefont {Pelucchi}\ \emph {et~al.}(2021)\citenamefont
  {Pelucchi}, \citenamefont {Fagas}, \citenamefont {Aharonovich}, \citenamefont
  {Englund}, \citenamefont {Figueroa}, \citenamefont {Gong}, \citenamefont
  {Hannes}, \citenamefont {Liu}, \citenamefont {Lu}, \citenamefont {Matsuda},
  \citenamefont {Pan}, \citenamefont {Schreck}, \citenamefont {Sciarrino},
  \citenamefont {Silberhorn}, \citenamefont {Wang},\ and\ \citenamefont
  {J\"{o}ns}}]{pelucchi2022potential}%
  \BibitemOpen
  \bibfield  {author} {\bibinfo {author} {\bibfnamefont {E.}~\bibnamefont
  {Pelucchi}}, \bibinfo {author} {\bibfnamefont {G.}~\bibnamefont {Fagas}},
  \bibinfo {author} {\bibfnamefont {I.}~\bibnamefont {Aharonovich}}, \bibinfo
  {author} {\bibfnamefont {D.}~\bibnamefont {Englund}}, \bibinfo {author}
  {\bibfnamefont {E.}~\bibnamefont {Figueroa}}, \bibinfo {author}
  {\bibfnamefont {Q.}~\bibnamefont {Gong}}, \bibinfo {author} {\bibfnamefont
  {H.}~\bibnamefont {Hannes}}, \bibinfo {author} {\bibfnamefont
  {J.}~\bibnamefont {Liu}}, \bibinfo {author} {\bibfnamefont {C.-Y.}\
  \bibnamefont {Lu}}, \bibinfo {author} {\bibfnamefont {N.}~\bibnamefont
  {Matsuda}}, \bibinfo {author} {\bibfnamefont {J.-W.}\ \bibnamefont {Pan}},
  \bibinfo {author} {\bibfnamefont {F.}~\bibnamefont {Schreck}}, \bibinfo
  {author} {\bibfnamefont {F.}~\bibnamefont {Sciarrino}}, \bibinfo {author}
  {\bibfnamefont {C.}~\bibnamefont {Silberhorn}}, \bibinfo {author}
  {\bibfnamefont {J.}~\bibnamefont {Wang}},\ and\ \bibinfo {author}
  {\bibfnamefont {K.~D.}\ \bibnamefont {J\"{o}ns}},\ }\bibfield  {title}
  {\bibinfo {title} {The potential and global outlook of integrated photonics
  for quantum technologies},\ }\href
  {https://doi.org/10.1038/s42254-021-00398-z} {\bibfield  {journal} {\bibinfo
  {journal} {Nature Reviews Physics}\ }\textbf {\bibinfo {volume} {4}},\
  \bibinfo {pages} {194–208} (\bibinfo {year} {2021})}\BibitemShut {NoStop}%
\bibitem [{\citenamefont {Wang}\ \emph {et~al.}(2019)\citenamefont {Wang},
  \citenamefont {Sciarrino}, \citenamefont {Laing},\ and\ \citenamefont
  {Thompson}}]{wang2020integrated}%
  \BibitemOpen
  \bibfield  {author} {\bibinfo {author} {\bibfnamefont {J.}~\bibnamefont
  {Wang}}, \bibinfo {author} {\bibfnamefont {F.}~\bibnamefont {Sciarrino}},
  \bibinfo {author} {\bibfnamefont {A.}~\bibnamefont {Laing}},\ and\ \bibinfo
  {author} {\bibfnamefont {M.~G.}\ \bibnamefont {Thompson}},\ }\bibfield
  {title} {\bibinfo {title} {Integrated photonic quantum technologies},\ }\href
  {https://doi.org/10.1038/s41566-019-0532-1} {\bibfield  {journal} {\bibinfo
  {journal} {Nature Photonics}\ }\textbf {\bibinfo {volume} {14}},\ \bibinfo
  {pages} {273–284} (\bibinfo {year} {2019})}\BibitemShut {NoStop}%
\bibitem [{\citenamefont {Giordani}\ \emph
  {et~al.}(2023{\natexlab{a}})\citenamefont {Giordani}, \citenamefont {Hoch},
  \citenamefont {Carvacho}, \citenamefont {Spagnolo},\ and\ \citenamefont
  {Sciarrino}}]{giordani2023integrated}%
  \BibitemOpen
  \bibfield  {author} {\bibinfo {author} {\bibfnamefont {T.}~\bibnamefont
  {Giordani}}, \bibinfo {author} {\bibfnamefont {F.}~\bibnamefont {Hoch}},
  \bibinfo {author} {\bibfnamefont {G.}~\bibnamefont {Carvacho}}, \bibinfo
  {author} {\bibfnamefont {N.}~\bibnamefont {Spagnolo}},\ and\ \bibinfo
  {author} {\bibfnamefont {F.}~\bibnamefont {Sciarrino}},\ }\bibfield  {title}
  {\bibinfo {title} {Integrated photonics in quantum technologies},\ }\href
  {https://doi.org/10.1007/s40766-023-00040-x} {\bibfield  {journal} {\bibinfo
  {journal} {La Rivista del Nuovo Cimento}\ }\textbf {\bibinfo {volume} {46}},\
  \bibinfo {pages} {71–103} (\bibinfo {year}
  {2023}{\natexlab{a}})}\BibitemShut {NoStop}%
\bibitem [{\citenamefont {Crespi}\ \emph {et~al.}(2013)\citenamefont {Crespi},
  \citenamefont {Osellame}, \citenamefont {Ramponi}, \citenamefont {Brod},
  \citenamefont {Galvão}, \citenamefont {Spagnolo}, \citenamefont {Vitelli},
  \citenamefont {Maiorino}, \citenamefont {Mataloni},\ and\ \citenamefont
  {Sciarrino}}]{crespi2013integrated}%
  \BibitemOpen
  \bibfield  {author} {\bibinfo {author} {\bibfnamefont {A.}~\bibnamefont
  {Crespi}}, \bibinfo {author} {\bibfnamefont {R.}~\bibnamefont {Osellame}},
  \bibinfo {author} {\bibfnamefont {R.}~\bibnamefont {Ramponi}}, \bibinfo
  {author} {\bibfnamefont {D.~J.}\ \bibnamefont {Brod}}, \bibinfo {author}
  {\bibfnamefont {E.~F.}\ \bibnamefont {Galvão}}, \bibinfo {author}
  {\bibfnamefont {N.}~\bibnamefont {Spagnolo}}, \bibinfo {author}
  {\bibfnamefont {C.}~\bibnamefont {Vitelli}}, \bibinfo {author} {\bibfnamefont
  {E.}~\bibnamefont {Maiorino}}, \bibinfo {author} {\bibfnamefont
  {P.}~\bibnamefont {Mataloni}},\ and\ \bibinfo {author} {\bibfnamefont
  {F.}~\bibnamefont {Sciarrino}},\ }\bibfield  {title} {\bibinfo {title}
  {Integrated multimode interferometers with arbitrary designs for photonic
  boson sampling},\ }\href {https://doi.org/10.1038/nphoton.2013.112}
  {\bibfield  {journal} {\bibinfo  {journal} {Nature Photonics}\ }\textbf
  {\bibinfo {volume} {7}},\ \bibinfo {pages} {545–549} (\bibinfo {year}
  {2013})}\BibitemShut {NoStop}%
\bibitem [{\citenamefont {Bennett}(1992)}]{Bennett1992}%
  \BibitemOpen
  \bibfield  {author} {\bibinfo {author} {\bibfnamefont {C.~H.}\ \bibnamefont
  {Bennett}},\ }\bibfield  {title} {\bibinfo {title} {Quantum cryptography
  using any two nonorthogonal states},\ }\href
  {https://doi.org/10.1103/physrevlett.68.3121} {\bibfield  {journal} {\bibinfo
   {journal} {Physical Review Letters}\ }\textbf {\bibinfo {volume} {68}},\
  \bibinfo {pages} {3121–3124} (\bibinfo {year} {1992})}\BibitemShut
  {NoStop}%
\bibitem [{\citenamefont {Luo}\ \emph {et~al.}(2023)\citenamefont {Luo},
  \citenamefont {Cao}, \citenamefont {Shi}, \citenamefont {Wan}, \citenamefont
  {Zhang}, \citenamefont {Li}, \citenamefont {Chen}, \citenamefont {Li},
  \citenamefont {Li}, \citenamefont {Wang}, \citenamefont {Sun}, \citenamefont
  {Karim}, \citenamefont {Cai}, \citenamefont {Kwek},\ and\ \citenamefont
  {Liu}}]{Luo2023}%
  \BibitemOpen
  \bibfield  {author} {\bibinfo {author} {\bibfnamefont {W.}~\bibnamefont
  {Luo}}, \bibinfo {author} {\bibfnamefont {L.}~\bibnamefont {Cao}}, \bibinfo
  {author} {\bibfnamefont {Y.}~\bibnamefont {Shi}}, \bibinfo {author}
  {\bibfnamefont {L.}~\bibnamefont {Wan}}, \bibinfo {author} {\bibfnamefont
  {H.}~\bibnamefont {Zhang}}, \bibinfo {author} {\bibfnamefont
  {S.}~\bibnamefont {Li}}, \bibinfo {author} {\bibfnamefont {G.}~\bibnamefont
  {Chen}}, \bibinfo {author} {\bibfnamefont {Y.}~\bibnamefont {Li}}, \bibinfo
  {author} {\bibfnamefont {S.}~\bibnamefont {Li}}, \bibinfo {author}
  {\bibfnamefont {Y.}~\bibnamefont {Wang}}, \bibinfo {author} {\bibfnamefont
  {S.}~\bibnamefont {Sun}}, \bibinfo {author} {\bibfnamefont {M.~F.}\
  \bibnamefont {Karim}}, \bibinfo {author} {\bibfnamefont {H.}~\bibnamefont
  {Cai}}, \bibinfo {author} {\bibfnamefont {L.~C.}\ \bibnamefont {Kwek}},\ and\
  \bibinfo {author} {\bibfnamefont {A.~Q.}\ \bibnamefont {Liu}},\ }\bibfield
  {title} {\bibinfo {title} {Recent progress in quantum photonic chips for
  quantum communication and internet},\ }\bibfield  {journal} {\bibinfo
  {journal} {Light: Science \& applications}\ }\textbf {\bibinfo {volume}
  {12}},\ \href {https://doi.org/10.1038/s41377-023-01173-8}
  {10.1038/s41377-023-01173-8} (\bibinfo {year} {2023})\BibitemShut {NoStop}%
\bibitem [{\citenamefont {Di~Matteo}(2014)}]{di2014short}%
  \BibitemOpen
  \bibfield  {author} {\bibinfo {author} {\bibfnamefont {O.}~\bibnamefont
  {Di~Matteo}},\ }\bibfield  {title} {\bibinfo {title} {A short introduction to
  unitary 2-designs},\ }\href@noop {} {\bibfield  {journal} {\bibinfo
  {journal} {CS867/QIC890}\ } (\bibinfo {year} {2014})}\BibitemShut {NoStop}%
\bibitem [{\citenamefont {Dankert}\ \emph {et~al.}(2009)\citenamefont
  {Dankert}, \citenamefont {Cleve}, \citenamefont {Emerson},\ and\
  \citenamefont {Livine}}]{dankert2009exact}%
  \BibitemOpen
  \bibfield  {author} {\bibinfo {author} {\bibfnamefont {C.}~\bibnamefont
  {Dankert}}, \bibinfo {author} {\bibfnamefont {R.}~\bibnamefont {Cleve}},
  \bibinfo {author} {\bibfnamefont {J.}~\bibnamefont {Emerson}},\ and\ \bibinfo
  {author} {\bibfnamefont {E.}~\bibnamefont {Livine}},\ }\bibfield  {title}
  {\bibinfo {title} {Exact and approximate unitary 2-designs and their
  application to fidelity estimation},\ }\href@noop {} {\bibfield  {journal}
  {\bibinfo  {journal} {Physical Review A}\ }\textbf {\bibinfo {volume} {80}},\
  \bibinfo {pages} {012304} (\bibinfo {year} {2009})}\BibitemShut {NoStop}%
\bibitem [{\citenamefont {Duan}\ and\ \citenamefont
  {Guo}(1998)}]{duan1998probabilistic}%
  \BibitemOpen
  \bibfield  {author} {\bibinfo {author} {\bibfnamefont {L.-M.}\ \bibnamefont
  {Duan}}\ and\ \bibinfo {author} {\bibfnamefont {G.-C.}\ \bibnamefont {Guo}},\
  }\bibfield  {title} {\bibinfo {title} {Probabilistic cloning and
  identification of linearly independent quantum states},\ }\href
  {https://doi.org/10.1103/physrevlett.80.4999} {\bibfield  {journal} {\bibinfo
   {journal} {Physical Review Letters}\ }\textbf {\bibinfo {volume} {80}},\
  \bibinfo {pages} {4999–5002} (\bibinfo {year} {1998})}\BibitemShut
  {NoStop}%
\bibitem [{\citenamefont {Cerf}\ \emph {et~al.}(2002)\citenamefont {Cerf},
  \citenamefont {Durt},\ and\ \citenamefont {Gisin}}]{Cerf2002}%
  \BibitemOpen
  \bibfield  {author} {\bibinfo {author} {\bibfnamefont {N.}~\bibnamefont
  {Cerf}}, \bibinfo {author} {\bibfnamefont {T.}~\bibnamefont {Durt}},\ and\
  \bibinfo {author} {\bibfnamefont {N.}~\bibnamefont {Gisin}},\ }\bibfield
  {title} {\bibinfo {title} {Cloning a qutrit},\ }\href
  {https://doi.org/10.1080/09500340110109043} {\bibfield  {journal} {\bibinfo
  {journal} {Journal of Modern Optics}\ }\textbf {\bibinfo {volume} {49}},\
  \bibinfo {pages} {1355–1373} (\bibinfo {year} {2002})}\BibitemShut
  {NoStop}%
\bibitem [{\citenamefont {Fan}(2003)}]{Fan2003}%
  \BibitemOpen
  \bibfield  {author} {\bibinfo {author} {\bibfnamefont {H.}~\bibnamefont
  {Fan}},\ }\bibfield  {title} {\bibinfo {title} {Quantum cloning of mixed
  states in symmetric subspaces},\ }\href
  {https://doi.org/10.1103/physreva.68.054301} {\bibfield  {journal} {\bibinfo
  {journal} {Physical Review A}\ }\textbf {\bibinfo {volume} {68}},\ \bibinfo
  {pages} {054301} (\bibinfo {year} {2003})}\BibitemShut {NoStop}%
\bibitem [{\citenamefont {Cerf}(2002)}]{Cerf2002a}%
  \BibitemOpen
  \bibfield  {author} {\bibinfo {author} {\bibfnamefont {N.}~\bibnamefont
  {Cerf}},\ }\bibfield  {title} {\bibinfo {title} {Quantum cloning with
  continuous variables},\ }in\ \href
  {https://doi.org/10.1007/978-94-015-1258-9_20} {\emph {\bibinfo {booktitle}
  {Quantum information with continuous variables}}},\ \bibinfo {editor} {edited
  by\ \bibinfo {editor} {\bibfnamefont {S.~L.}\ \bibnamefont {Braunstein}}\
  and\ \bibinfo {editor} {\bibfnamefont {A.~K.}\ \bibnamefont {Pati}}}\
  (\bibinfo  {publisher} {Springer Dordrecht},\ \bibinfo {year} {2002})\ pp.\
  \bibinfo {pages} {277--293}\BibitemShut {NoStop}%
\bibitem [{\citenamefont {Zhang}\ \emph {et~al.}(2010)\citenamefont {Zhang},
  \citenamefont {Dai}, \citenamefont {Cao},\ and\ \citenamefont
  {Yang}}]{Zhang2010}%
  \BibitemOpen
  \bibfield  {author} {\bibinfo {author} {\bibfnamefont {W.-H.}\ \bibnamefont
  {Zhang}}, \bibinfo {author} {\bibfnamefont {J.-L.}\ \bibnamefont {Dai}},
  \bibinfo {author} {\bibfnamefont {Z.-L.}\ \bibnamefont {Cao}},\ and\ \bibinfo
  {author} {\bibfnamefont {M.}~\bibnamefont {Yang}},\ }\bibfield  {title}
  {\bibinfo {title} {Probabilistically perfect quantum cloning and unambiguous
  state discrimination},\ }\href {https://doi.org/10.1016/j.optcom.2010.05.046}
  {\bibfield  {journal} {\bibinfo  {journal} {Optics Communications}\ }\textbf
  {\bibinfo {volume} {283}},\ \bibinfo {pages} {3818–3824} (\bibinfo {year}
  {2010})}\BibitemShut {NoStop}%
\bibitem [{\citenamefont {Fan}\ \emph {et~al.}(2001)\citenamefont {Fan},
  \citenamefont {Matsumoto}, \citenamefont {Wang},\ and\ \citenamefont
  {Wadati}}]{Fan2001}%
  \BibitemOpen
  \bibfield  {author} {\bibinfo {author} {\bibfnamefont {H.}~\bibnamefont
  {Fan}}, \bibinfo {author} {\bibfnamefont {K.}~\bibnamefont {Matsumoto}},
  \bibinfo {author} {\bibfnamefont {X.-B.}\ \bibnamefont {Wang}},\ and\
  \bibinfo {author} {\bibfnamefont {M.}~\bibnamefont {Wadati}},\ }\bibfield
  {title} {\bibinfo {title} {Quantum cloning machines for equatorial qubits},\
  }\href {https://doi.org/10.1103/physreva.65.012304} {\bibfield  {journal}
  {\bibinfo  {journal} {Physical Review A}\ }\textbf {\bibinfo {volume} {65}},\
  \bibinfo {pages} {012304} (\bibinfo {year} {2001})}\BibitemShut {NoStop}%
\bibitem [{\citenamefont {Hashagen}(2017)}]{Hashagen2017}%
  \BibitemOpen
  \bibfield  {author} {\bibinfo {author} {\bibfnamefont {A.-L.}\ \bibnamefont
  {Hashagen}},\ }\bibfield  {title} {\bibinfo {title} {Universal asymmetric
  quantum cloning revisited},\ }\href {https://doi.org/10.26421/qic17.9-10-2}
  {\bibfield  {journal} {\bibinfo  {journal} {Quantum Information and
  Computation}\ }\textbf {\bibinfo {volume} {17}},\ \bibinfo {pages}
  {747–778} (\bibinfo {year} {2017})}\BibitemShut {NoStop}%
\bibitem [{\citenamefont {Pentangelo}\ \emph {et~al.}(2024)\citenamefont
  {Pentangelo}, \citenamefont {Di~Giano}, \citenamefont {Piacentini},
  \citenamefont {Arpe}, \citenamefont {Ceccarelli}, \citenamefont {Crespi},\
  and\ \citenamefont {Osellame}}]{Pentangelo2024}%
  \BibitemOpen
  \bibfield  {author} {\bibinfo {author} {\bibfnamefont {C.}~\bibnamefont
  {Pentangelo}}, \bibinfo {author} {\bibfnamefont {N.}~\bibnamefont
  {Di~Giano}}, \bibinfo {author} {\bibfnamefont {S.}~\bibnamefont
  {Piacentini}}, \bibinfo {author} {\bibfnamefont {R.}~\bibnamefont {Arpe}},
  \bibinfo {author} {\bibfnamefont {F.}~\bibnamefont {Ceccarelli}}, \bibinfo
  {author} {\bibfnamefont {A.}~\bibnamefont {Crespi}},\ and\ \bibinfo {author}
  {\bibfnamefont {R.}~\bibnamefont {Osellame}},\ }\bibfield  {title} {\bibinfo
  {title} {High-fidelity and polarization-insensitive universal photonic
  processors fabricated by femtosecond laser writing},\ }\href
  {https://doi.org/10.1515/nanoph-2023-0636} {\bibfield  {journal} {\bibinfo
  {journal} {Nanophotonics}\ }\textbf {\bibinfo {volume} {13}},\ \bibinfo
  {pages} {2259–2270} (\bibinfo {year} {2024})}\BibitemShut {NoStop}%
\bibitem [{\citenamefont {Giordani}\ \emph
  {et~al.}(2023{\natexlab{b}})\citenamefont {Giordani}, \citenamefont {Wagner},
  \citenamefont {Esposito}, \citenamefont {Camillini}, \citenamefont {Hoch},
  \citenamefont {Carvacho}, \citenamefont {Pentangelo}, \citenamefont
  {Ceccarelli}, \citenamefont {Piacentini}, \citenamefont {Crespi},
  \citenamefont {Spagnolo}, \citenamefont {Osellame}, \citenamefont
  {Galv\~{a}o},\ and\ \citenamefont {Sciarrino}}]{giordani2023experimental}%
  \BibitemOpen
  \bibfield  {author} {\bibinfo {author} {\bibfnamefont {T.}~\bibnamefont
  {Giordani}}, \bibinfo {author} {\bibfnamefont {R.}~\bibnamefont {Wagner}},
  \bibinfo {author} {\bibfnamefont {C.}~\bibnamefont {Esposito}}, \bibinfo
  {author} {\bibfnamefont {A.}~\bibnamefont {Camillini}}, \bibinfo {author}
  {\bibfnamefont {F.}~\bibnamefont {Hoch}}, \bibinfo {author} {\bibfnamefont
  {G.}~\bibnamefont {Carvacho}}, \bibinfo {author} {\bibfnamefont
  {C.}~\bibnamefont {Pentangelo}}, \bibinfo {author} {\bibfnamefont
  {F.}~\bibnamefont {Ceccarelli}}, \bibinfo {author} {\bibfnamefont
  {S.}~\bibnamefont {Piacentini}}, \bibinfo {author} {\bibfnamefont
  {A.}~\bibnamefont {Crespi}}, \bibinfo {author} {\bibfnamefont
  {N.}~\bibnamefont {Spagnolo}}, \bibinfo {author} {\bibfnamefont
  {R.}~\bibnamefont {Osellame}}, \bibinfo {author} {\bibfnamefont {E.~F.}\
  \bibnamefont {Galv\~{a}o}},\ and\ \bibinfo {author} {\bibfnamefont
  {F.}~\bibnamefont {Sciarrino}},\ }\bibfield  {title} {\bibinfo {title}
  {Experimental certification of contextuality, coherence, and dimension in a
  programmable universal photonic processor},\ }\href
  {https://doi.org/10.1126/sciadv.adj4249} {\bibfield  {journal} {\bibinfo
  {journal} {Science Advances}\ }\textbf {\bibinfo {volume} {9}},\ \bibinfo
  {pages} {eadj4249} (\bibinfo {year} {2023}{\natexlab{b}})}\BibitemShut
  {NoStop}%
\bibitem [{\citenamefont {Corrielli}\ \emph {et~al.}(2021)\citenamefont
  {Corrielli}, \citenamefont {Crespi},\ and\ \citenamefont
  {Osellame}}]{Corrielli2021}%
  \BibitemOpen
  \bibfield  {author} {\bibinfo {author} {\bibfnamefont {G.}~\bibnamefont
  {Corrielli}}, \bibinfo {author} {\bibfnamefont {A.}~\bibnamefont {Crespi}},\
  and\ \bibinfo {author} {\bibfnamefont {R.}~\bibnamefont {Osellame}},\
  }\bibfield  {title} {\bibinfo {title} {Femtosecond laser micromachining for
  integrated quantum photonics},\ }\href
  {https://doi.org/10.1515/nanoph-2021-0419} {\bibfield  {journal} {\bibinfo
  {journal} {Nanophotonics}\ }\textbf {\bibinfo {volume} {10}},\ \bibinfo
  {pages} {3789–3812} (\bibinfo {year} {2021})}\BibitemShut {NoStop}%
\bibitem [{\citenamefont {Ceccarelli}\ \emph {et~al.}(2020)\citenamefont
  {Ceccarelli}, \citenamefont {Atzeni}, \citenamefont {Pentangelo},
  \citenamefont {Pellegatta}, \citenamefont {Crespi},\ and\ \citenamefont
  {Osellame}}]{Ceccarelli2020}%
  \BibitemOpen
  \bibfield  {author} {\bibinfo {author} {\bibfnamefont {F.}~\bibnamefont
  {Ceccarelli}}, \bibinfo {author} {\bibfnamefont {S.}~\bibnamefont {Atzeni}},
  \bibinfo {author} {\bibfnamefont {C.}~\bibnamefont {Pentangelo}}, \bibinfo
  {author} {\bibfnamefont {F.}~\bibnamefont {Pellegatta}}, \bibinfo {author}
  {\bibfnamefont {A.}~\bibnamefont {Crespi}},\ and\ \bibinfo {author}
  {\bibfnamefont {R.}~\bibnamefont {Osellame}},\ }\bibfield  {title} {\bibinfo
  {title} {Low power reconfigurability and reduced crosstalk in integrated
  photonic circuits fabricated by femtosecond laser micromachining},\
  }\bibfield  {journal} {\bibinfo  {journal} {Laser amp; Photonics Reviews}\
  }\textbf {\bibinfo {volume} {14}},\ \href
  {https://doi.org/10.1002/lpor.202000024} {10.1002/lpor.202000024} (\bibinfo
  {year} {2020})\BibitemShut {NoStop}%
\bibitem [{\citenamefont {Clements}\ \emph {et~al.}(2016)\citenamefont
  {Clements}, \citenamefont {Humphreys}, \citenamefont {Metcalf}, \citenamefont
  {Kolthammer},\ and\ \citenamefont {Walsmley}}]{Clements2016}%
  \BibitemOpen
  \bibfield  {author} {\bibinfo {author} {\bibfnamefont {W.~R.}\ \bibnamefont
  {Clements}}, \bibinfo {author} {\bibfnamefont {P.~C.}\ \bibnamefont
  {Humphreys}}, \bibinfo {author} {\bibfnamefont {B.~J.}\ \bibnamefont
  {Metcalf}}, \bibinfo {author} {\bibfnamefont {W.~S.}\ \bibnamefont
  {Kolthammer}},\ and\ \bibinfo {author} {\bibfnamefont {I.~A.}\ \bibnamefont
  {Walsmley}},\ }\bibfield  {title} {\bibinfo {title} {Optimal design for
  universal multiport interferometers},\ }\href
  {https://doi.org/10.1364/optica.3.001460} {\bibfield  {journal} {\bibinfo
  {journal} {Optica}\ }\textbf {\bibinfo {volume} {3}},\ \bibinfo {pages}
  {1460} (\bibinfo {year} {2016})}\BibitemShut {NoStop}%
\bibitem [{\citenamefont {Flamini}\ \emph {et~al.}(2015)\citenamefont
  {Flamini}, \citenamefont {Magrini}, \citenamefont {Rab}, \citenamefont
  {Spagnolo}, \citenamefont {D'ambrosio}, \citenamefont {Mataloni},
  \citenamefont {Sciarrino}, \citenamefont {Zandrini}, \citenamefont {Crespi},
  \citenamefont {Ramponi},\ and\ \citenamefont
  {Osellame}}]{flamini2015thermally}%
  \BibitemOpen
  \bibfield  {author} {\bibinfo {author} {\bibfnamefont {F.}~\bibnamefont
  {Flamini}}, \bibinfo {author} {\bibfnamefont {L.}~\bibnamefont {Magrini}},
  \bibinfo {author} {\bibfnamefont {A.~S.}\ \bibnamefont {Rab}}, \bibinfo
  {author} {\bibfnamefont {N.}~\bibnamefont {Spagnolo}}, \bibinfo {author}
  {\bibfnamefont {V.}~\bibnamefont {D'ambrosio}}, \bibinfo {author}
  {\bibfnamefont {P.}~\bibnamefont {Mataloni}}, \bibinfo {author}
  {\bibfnamefont {F.}~\bibnamefont {Sciarrino}}, \bibinfo {author}
  {\bibfnamefont {T.}~\bibnamefont {Zandrini}}, \bibinfo {author}
  {\bibfnamefont {A.}~\bibnamefont {Crespi}}, \bibinfo {author} {\bibfnamefont
  {R.}~\bibnamefont {Ramponi}},\ and\ \bibinfo {author} {\bibfnamefont
  {R.}~\bibnamefont {Osellame}},\ }\bibfield  {title} {\bibinfo {title}
  {Thermally reconfigurable quantum photonic circuits at telecom wavelength by
  femtosecond laser micromachining},\ }\href
  {https://doi.org/10.1038/lsa.2015.127} {\bibfield  {journal} {\bibinfo
  {journal} {Light: Science \& Applications}\ }\textbf {\bibinfo {volume}
  {4}},\ \bibinfo {pages} {e354} (\bibinfo {year} {2015})}\BibitemShut
  {NoStop}%
\bibitem [{\citenamefont {Nelder}\ and\ \citenamefont
  {Mead}(1965)}]{nelder1965simplex}%
  \BibitemOpen
  \bibfield  {author} {\bibinfo {author} {\bibfnamefont {J.~A.}\ \bibnamefont
  {Nelder}}\ and\ \bibinfo {author} {\bibfnamefont {R.}~\bibnamefont {Mead}},\
  }\bibfield  {title} {\bibinfo {title} {A simplex method for function
  minimization},\ }\href {https://doi.org/10.1093/comjnl/7.4.308} {\bibfield
  {journal} {\bibinfo  {journal} {The computer journal}\ }\textbf {\bibinfo
  {volume} {7}},\ \bibinfo {pages} {308} (\bibinfo {year} {1965})}\BibitemShut
  {NoStop}%
\bibitem [{\citenamefont {Poderini}\ \emph {et~al.}(2022)\citenamefont
  {Poderini}, \citenamefont {Polino}, \citenamefont {Rodari}, \citenamefont
  {Suprano}, \citenamefont {Chaves},\ and\ \citenamefont
  {Sciarrino}}]{poderini2022ab}%
  \BibitemOpen
  \bibfield  {author} {\bibinfo {author} {\bibfnamefont {D.}~\bibnamefont
  {Poderini}}, \bibinfo {author} {\bibfnamefont {E.}~\bibnamefont {Polino}},
  \bibinfo {author} {\bibfnamefont {G.}~\bibnamefont {Rodari}}, \bibinfo
  {author} {\bibfnamefont {A.}~\bibnamefont {Suprano}}, \bibinfo {author}
  {\bibfnamefont {R.}~\bibnamefont {Chaves}},\ and\ \bibinfo {author}
  {\bibfnamefont {F.}~\bibnamefont {Sciarrino}},\ }\bibfield  {title} {\bibinfo
  {title} {Ab initio experimental violation of bell inequalities},\ }\href
  {https://doi.org/10.1103/PhysRevResearch.4.013159} {\bibfield  {journal}
  {\bibinfo  {journal} {Physical Review Research}\ }\textbf {\bibinfo {volume}
  {4}},\ \bibinfo {pages} {013159} (\bibinfo {year} {2022})}\BibitemShut
  {NoStop}%
\bibitem [{\citenamefont {Huang}(2018)}]{Huang2018}%
  \BibitemOpen
  \bibfield  {author} {\bibinfo {author} {\bibfnamefont {X.}~\bibnamefont
  {Huang}},\ }\bibfield  {title} {\bibinfo {title} {Robust simplex algorithm
  for online optimization},\ }\href
  {https://doi.org/10.1103/physrevaccelbeams.21.104601} {\bibfield  {journal}
  {\bibinfo  {journal} {Physical Review Accelerators and Beams}\ }\textbf
  {\bibinfo {volume} {21}},\ \bibinfo {pages} {104601} (\bibinfo {year}
  {2018})}\BibitemShut {NoStop}%
\bibitem [{\citenamefont {Heurtel}\ \emph {et~al.}(2023)\citenamefont
  {Heurtel}, \citenamefont {Fyrillas}, \citenamefont {Gliniasty}, \citenamefont
  {Le~Bihan}, \citenamefont {Malherbe}, \citenamefont {Pailhas}, \citenamefont
  {Bertasi}, \citenamefont {Bourdoncle}, \citenamefont {Emeriau}, \citenamefont
  {Mezher}, \citenamefont {Music}, \citenamefont {Belabas}, \citenamefont
  {Valiron}, \citenamefont {Senellart}, \citenamefont {Mansfield},\ and\
  \citenamefont {Senellart}}]{heurtel2023perceval}%
  \BibitemOpen
  \bibfield  {author} {\bibinfo {author} {\bibfnamefont {N.}~\bibnamefont
  {Heurtel}}, \bibinfo {author} {\bibfnamefont {A.}~\bibnamefont {Fyrillas}},
  \bibinfo {author} {\bibfnamefont {G.~d.}\ \bibnamefont {Gliniasty}}, \bibinfo
  {author} {\bibfnamefont {R.}~\bibnamefont {Le~Bihan}}, \bibinfo {author}
  {\bibfnamefont {S.}~\bibnamefont {Malherbe}}, \bibinfo {author}
  {\bibfnamefont {M.}~\bibnamefont {Pailhas}}, \bibinfo {author} {\bibfnamefont
  {E.}~\bibnamefont {Bertasi}}, \bibinfo {author} {\bibfnamefont
  {B.}~\bibnamefont {Bourdoncle}}, \bibinfo {author} {\bibfnamefont {P.-E.}\
  \bibnamefont {Emeriau}}, \bibinfo {author} {\bibfnamefont {R.}~\bibnamefont
  {Mezher}}, \bibinfo {author} {\bibfnamefont {L.}~\bibnamefont {Music}},
  \bibinfo {author} {\bibfnamefont {N.}~\bibnamefont {Belabas}}, \bibinfo
  {author} {\bibfnamefont {B.}~\bibnamefont {Valiron}}, \bibinfo {author}
  {\bibfnamefont {P.}~\bibnamefont {Senellart}}, \bibinfo {author}
  {\bibfnamefont {S.}~\bibnamefont {Mansfield}},\ and\ \bibinfo {author}
  {\bibfnamefont {J.}~\bibnamefont {Senellart}},\ }\bibfield  {title} {\bibinfo
  {title} {Perceval: {A} {S}oftware {P}latform for {D}iscrete {V}ariable
  {P}hotonic {Q}uantum {C}omputing},\ }\href
  {https://doi.org/10.22331/q-2023-02-21-931} {\bibfield  {journal} {\bibinfo
  {journal} {{Quantum}}\ }\textbf {\bibinfo {volume} {7}},\ \bibinfo {pages}
  {931} (\bibinfo {year} {2023})}\BibitemShut {NoStop}%
\bibitem [{\citenamefont {Wang}\ \emph {et~al.}(2016)\citenamefont {Wang},
  \citenamefont {Bonneau}, \citenamefont {Villa}, \citenamefont {Silverstone},
  \citenamefont {Santagati}, \citenamefont {Miki}, \citenamefont {Yamashita},
  \citenamefont {Fujiwara}, \citenamefont {Sasaki}, \citenamefont {Terai},
  \citenamefont {Tanner}, \citenamefont {Natarajan}, \citenamefont {Hadfield},
  \citenamefont {O'Brien},\ and\ \citenamefont {Thompson}}]{Wang:16}%
  \BibitemOpen
  \bibfield  {author} {\bibinfo {author} {\bibfnamefont {J.}~\bibnamefont
  {Wang}}, \bibinfo {author} {\bibfnamefont {D.}~\bibnamefont {Bonneau}},
  \bibinfo {author} {\bibfnamefont {M.}~\bibnamefont {Villa}}, \bibinfo
  {author} {\bibfnamefont {J.~W.}\ \bibnamefont {Silverstone}}, \bibinfo
  {author} {\bibfnamefont {R.}~\bibnamefont {Santagati}}, \bibinfo {author}
  {\bibfnamefont {S.}~\bibnamefont {Miki}}, \bibinfo {author} {\bibfnamefont
  {T.}~\bibnamefont {Yamashita}}, \bibinfo {author} {\bibfnamefont
  {M.}~\bibnamefont {Fujiwara}}, \bibinfo {author} {\bibfnamefont
  {M.}~\bibnamefont {Sasaki}}, \bibinfo {author} {\bibfnamefont
  {H.}~\bibnamefont {Terai}}, \bibinfo {author} {\bibfnamefont {M.~G.}\
  \bibnamefont {Tanner}}, \bibinfo {author} {\bibfnamefont {C.~M.}\
  \bibnamefont {Natarajan}}, \bibinfo {author} {\bibfnamefont {R.~H.}\
  \bibnamefont {Hadfield}}, \bibinfo {author} {\bibfnamefont {J.~L.}\
  \bibnamefont {O'Brien}},\ and\ \bibinfo {author} {\bibfnamefont {M.~G.}\
  \bibnamefont {Thompson}},\ }\bibfield  {title} {\bibinfo {title}
  {Chip-to-chip quantum photonic interconnect by path-polarization
  interconversion},\ }\href {https://doi.org/10.1364/OPTICA.3.000407}
  {\bibfield  {journal} {\bibinfo  {journal} {Optica}\ }\textbf {\bibinfo
  {volume} {3}},\ \bibinfo {pages} {407} (\bibinfo {year} {2016})}\BibitemShut
  {NoStop}%
\bibitem [{\citenamefont {Ambainis}\ and\ \citenamefont
  {Emerson}(2007)}]{Ambainis2007}%
  \BibitemOpen
  \bibfield  {author} {\bibinfo {author} {\bibfnamefont {A.}~\bibnamefont
  {Ambainis}}\ and\ \bibinfo {author} {\bibfnamefont {J.}~\bibnamefont
  {Emerson}},\ }\bibfield  {title} {\bibinfo {title} {Quantum t-designs: t-wise
  independence in the quantum world},\ }in\ \href
  {https://doi.org/10.1109/CCC.2007.26} {\emph {\bibinfo {booktitle} {CCC '07:
  Proceedings of the Twenty-Second Annual IEEE Conference on Computational
  Complexity}}}\ (\bibinfo  {publisher} {ACM},\ \bibinfo {year}
  {2007})\BibitemShut {NoStop}%
\end{thebibliography}

%

\section*{Acknowledgements}

This work is supported by the European Union’s Horizon 2020 research and innovation program through the FET project PHOQUSING (“PHOtonic Quantum SamplING machine” - Grant Agreement No. 899544), and by ICSC – Centro Nazionale di Ricerca in High Performance Computing, Big Data and Quantum Computing, funded by European Union – NextGenerationEU. E.G acknowledges support from FCT – Fundaç\~{a}o para a Ciência e a Tecnologia (Portugal) via project CEECINST/00062/2018. The universal integrated interferometer was partially fabricated at PoliFAB, the micro- and nanofabrication facility of Politecnico di Milano (www.polifab.polimi.it/). C.P., F.C., and R.O. wish to thank the PoliFAB staff for their valuable technical support.

\section*{Disclosures} F.C. and R.O. are co-founders of the company Ephos. The other authors declare no competing interests.

{\section*{Data Availability Statement} Data underlying the results presented in this paper are available from the authors upon reasonable request.}

\end{document}



\title{Supplementary Materials: Variational quantum cloning machine on an integrated photonic interferometer}

\author{Francesco Hoch}
\affiliation{Dipartimento di Fisica - Sapienza Universit\`{a} di Roma, P.le Aldo Moro 5, I-00185 Roma, Italy}

\author{Giovanni Rodari}
\affiliation{Dipartimento di Fisica - Sapienza Universit\`{a} di Roma, P.le Aldo Moro 5, I-00185 Roma, Italy}

\author{Eugenio Caruccio}
\affiliation{Dipartimento di Fisica - Sapienza Universit\`{a} di Roma, P.le Aldo Moro 5, I-00185 Roma, Italy}

\author{Beatrice Polacchi}
\affiliation{Dipartimento di Fisica - Sapienza Universit\`{a} di Roma, P.le Aldo Moro 5, I-00185 Roma, Italy}

\author{Gonzalo Carvacho}
\affiliation{Dipartimento di Fisica - Sapienza Universit\`{a} di Roma, P.le Aldo Moro 5, I-00185 Roma, Italy}

\author{Taira Giordani}
\email[Corresponding author: ]{taira.giordani@uniroma1.it}
\affiliation{Dipartimento di Fisica - Sapienza Universit\`{a} di Roma, P.le Aldo Moro 5, I-00185 Roma, Italy}

\author{Mina Doosti}\affiliation{School of Informatics, University of Edinburgh, 10 Crichton Street, EH8 9AB Edinburgh, United Kingdom}

\author{Sebasti\`{a} Nicolau}
\affiliation{CNRS, LIP6, Sorbonne Université, 4 place Jussieu, 75005 Paris, France}
\affiliation{Escola T\`{e}cnica Superior d’Enginyeria de Telecomunicaci\'{o} de Barcelona, Universitat Polit\`{e}cnica de Catalunya, Carrer de Jordi Girona, 1-3, 08034 Barcelona, Spain}

\author{Ciro Pentangelo}
\affiliation{Istituto di Fotonica e Nanotecnologie, Consiglio Nazionale delle Ricerche (IFN-CNR),
Piazza Leonardo da Vinci, 32, I-20133 Milano, Italy}

\author{Simone Piacentini}
\affiliation{Istituto di Fotonica e Nanotecnologie, Consiglio Nazionale delle Ricerche (IFN-CNR), 
Piazza Leonardo da Vinci, 32, I-20133 Milano, Italy}

\author{Andrea Crespi}
\affiliation{Istituto di Fotonica e Nanotecnologie, Consiglio Nazionale delle Ricerche (IFN-CNR), 
Piazza Leonardo da Vinci, 32, I-20133 Milano, Italy}
\affiliation{Dipartimento di Fisica, Politecnico di Milano, Piazza Leonardo da Vinci 32, I-20133 Milano, Italy}

\author{Francesco Ceccarelli}
\affiliation{Istituto di Fotonica e Nanotecnologie, Consiglio Nazionale delle Ricerche (IFN-CNR), 
Piazza Leonardo da Vinci, 32, I-20133 Milano, Italy}

\author{Roberto Osellame}
\affiliation{Istituto di Fotonica e Nanotecnologie, Consiglio Nazionale delle Ricerche (IFN-CNR), 
Piazza Leonardo da Vinci, 32, I-20133 Milano, Italy}

\author{Ernesto F. Galv\~{a}o}
\affiliation{International Iberian Nanotechnology Laboratory (INL), Av. Mestre José Veiga, 4715-330 Braga, Portugal}
\affiliation{Instituto de F\'isica, Universidade Federal Fluminense, Av. Gal. Milton Tavares de Souza s/n, Niter\'oi, RJ, 24210-340, Brazil}

\author{Nicol\`o Spagnolo}
\affiliation{Dipartimento di Fisica - Sapienza Universit\`{a} di Roma, P.le Aldo Moro 5, I-00185 Roma, Italy}

\author{Fabio Sciarrino}
\affiliation{Dipartimento di Fisica - Sapienza Universit\`{a} di Roma, P.le Aldo Moro 5, I-00185 Roma, Italy}

\maketitle

{
\section{Experimental details}
\subsection{Dual rail encoding}
The experiment works with the photonic dual-rail encoding. This consists of a photon in a superposition of two optical modes.
To prepare photonic qubits, we use the scheme depicted in the preparation stage of Fig.~2b of the main text.
It consists of a Mach-Zehnder interferometer with a variable phase $\theta$ followed by a phase shifter $\phi$. If a photon is injected into the upper arm, the generated state is:
\begin{equation}
    \ket{\psi} = \sin \frac{\theta}{2}\ket{0}+\cos\frac{\theta}{2}e^{i\phi}\ket{1}
\end{equation}
up to a global phase.
}

{
To measure a state, we use a similar interferometer as depicted in the measurement stage of Fig.~2b that consists of a Phase shifter followed by a  Mach-Zehnder interferometer. This apparatus operates the projection on the basis $\{\ket{\xi }, \ket{\xi^\perp }\}$, where $\ket{\xi}$ is a target state of the training set. In particular, if at the input of the interferometer we have a dual-rail state $\ket{\psi}$, then the probability of detecting the photon in the first mode is
\begin{equation}
    P_0 = \abs{\braket{\xi}{\psi}}^2 \qquad \ket{\xi} = \sin \frac{\theta}{2}\ket{0}+\cos\frac{\theta}{2}e^{i\phi}\ket{1}
\end{equation}
which corresponds to the fidelity between the input state $\ket{\psi}$ and a state $\ket{\xi}$ that depends on the parameter of the beam splitter and the phase shifter.
With those two interferometers, we have all the elements to prepare and measure a dual rail qubit with the advantage that we do not need to perform a tomography of the state, but we can directly measure the state fidelity compared to the target state.
}

{\subsection{Calibration and control of the universal integrated device}
The 6-mode integrated chip was designed according to the universal layout proposed in Ref. \cite{Clements2016}. The device is equipped with 30 heaters that allow for full control over the phases of the interferometer. More precisely, the heaters, which are placed and patterned on the surface of the glass sample, induce a set of phases on the waveguides of the photon circuit through the so-called thermo-optic effect. Phases $\vec{\phi}$ are linearly related to the powers dissipated $\vec{P}$ by thermo-optic phase shifters, which operate as resistive elements when a current is injected: $\vec{\phi} = \vec{\phi}_0+\hat{A}\vec{P}$, where $\vec{\phi}_0$ is the vector of accumulated phases as a result of waveguide geometry when current is not supplied and $\hat{A}$ is the matrix of cross-talk coefficients that map the impact of phase shifters along the entire chip. The purpose of fully reconfigurable circuit calibration is to derive all the parameters contained in $\vec{\phi}_0$ and $\hat{A}$, also considering imperfections in manufacturing the directional couplers that constitute the Mach-Zehnder interferometer unit cell. Calibration parameters were derived by measuring interference fringes with a suitable procedure and were manipulated to implement a generic $6$x$6$ unitary transformation. At the end of the calibration procedure, the method was benchmarked by measuring the average fidelity on 1000 Haar random matrices: $\mathcal{F}=0.997\pm0.002$. As sources of noise, one must consider that the calibration model depends on a large number ($\sim O(10^2)$) of parameters obtained experimentally with different chip configurations. Therefore, the implemented unitary will consider all these experimental error spreads. Besides, the directional coupler defects inherently constrain the device's capabilities in constructing unitary matrices. Nevertheless, the harnessed calibration method provides a reliable and accurate way to work with the photonic integrated circuit. This is also demonstrated by the high value obtained with Haar random matrices.}

\begin{figure}[ht!]
\centering
\includegraphics[width=0.55\textwidth]{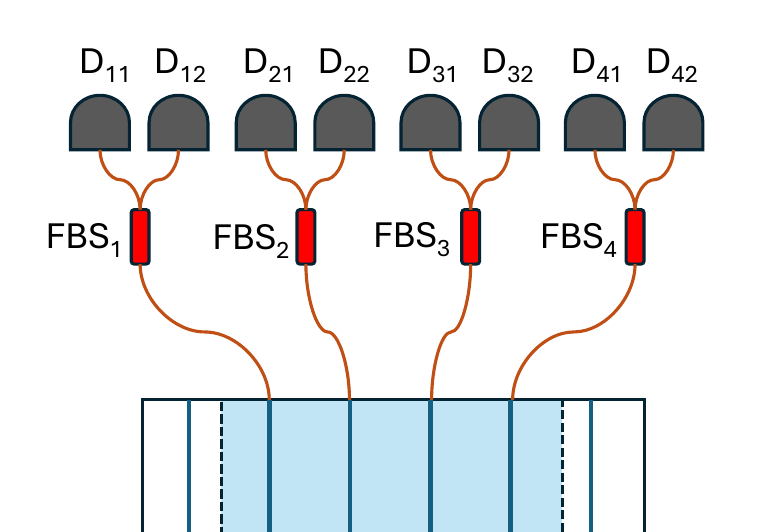}
\caption{{\textbf{Detection apparatus.} Photon number resolution is obtained probabilistically via the detection apparatus shown schematically in this figure. On each output mode $i$, a fiber beam splitter (FBS) is placed, and the two output ports are connected to two threshold detectors $D_{i1}$ and $D_{i2}$.}}
\label{fig:pnr}
\end{figure}

{\subsection{Detection method for photon-number resolution}
We provide below a detailed analysis of the detection method employed to obtain photon-number resolution up to $n=2$. The adopted single-photon detectors are avalanche photodiodes acting as threshold detectors, and thus providing clicks without the capability to resolve the number of impinging photons. However, estimation of some quantities, for instance, the success probabilities in Eq. (9), requires the capability to resolve the number of output photons on each mode. This can be obtained with threshold detectors by using a probabilistic approach \cite{rodari2024semi,rodari2024experimental}. On each output mode of the interferometer, we inserted a symmetric 50:50 fiber beam-splitter, where the two output ports are connected to two distinct threshold detectors (see Fig. \ref{fig:pnr}). When two photons are present in one output mode $i$ of the interferometer, with probability $1/2$ they exit from two distinct output ports of fiber beam splitter $\mathrm{FBS}_{i}$. Detecting a coincidence between the two detectors $\mathrm{D}_{i1}$ and $\mathrm{D}_{i2})$ positioned in these two output ports thus corresponds to selecting those events with two output photons from mode $i$ of the interferometer. This is obtained, as said, with probability $1/2$ acting as an additional effective detection efficiency for bunching terms. Conversely, if one photon is present in mode $i$ and one in mode $j$, the event is measured by detecting a coincidence between one of the two detectors $(\mathrm{D}_{i1}, \mathrm{D}_{i2})$ and one of the two detectors $(\mathrm{D}_{j1}, \mathrm{D}_{j2})$. In the actual experiment, each detector is characterized by a different detection efficiency ($\eta_{ij}$, being $i$ the output mode and $j=1,2$ the index labelling the output port of the fiber beam splitters). Furthermore, the fiber beam splitters are characterized by effective reflectivities $R_{i}$. These parameters have been measured before the experiment with independent measurements. This permitted to compute the effective relative overall efficiencies for the different output combinations, and thus calibrating the detection apparatus.}

{More specifically, let us call $p(n_{1}, n_{2}, n_{3}, n_{4})$ the probability at the output of the interferometer, before detection, where the $\{n_i\}$ are the number of photons in output modes $i$, where in our case $\sum_i n_i = 2$. The apparatus described above adds an effective efficiency $\eta_{\mathrm{eff}}(n_{1}, n_{2}, n_{3}, n_{4})$ for each output configuration, and thus the detected probabilities are:
\begin{equation}
p_{\mathrm{det}}(n_{1}, n_{2}, n_{3}, n_{4}) = \frac{p(n_{1}, n_{2}, n_{3}, n_{4}) \eta_{\mathrm{eff}}(n_{1}, n_{2}, n_{3}, n_{4})}{\sum_{n_{1}, n_{2}, n_{3}, n_{4}} p(n_{1}, n_{2}, n_{3}, n_{4}) \eta_{\mathrm{eff}}(n_{1}, n_{2}, n_{3}, n_{4})}.
\end{equation}
The parameters are due to the probabilistic splitting and depend on parameters $R_{i}$ and $\eta_{ij}$, according to:
\begin{gather}
   \eta_{\mathrm{eff}}(n_{1}, n_{2}, n_{3}, n_{4}) = \prod_{i=1}^{4} \tilde{\eta}_{i}(n_{i}), \\
\tilde{\eta}_{i}(n_{i}) = (n_i)! \sum_{\sigma \in C_{n_{i},n_{T}}} \prod_{k=1}^{n_i} Q_{i,\sigma_{k}}\eta_{i,\sigma_{k}},
\end{gather}
where $C_{n_i, n_T}$ is the set of combinations without replacement of $n_i$ elements in $n_T$ slots ( in our case $n_T = 2)$, while $Q_{i1} = R_{i}$ and $Q_{i2} = 1-R_{i}$. Thus, by an appropriate calibration it is possible to retrieve the probabilities $p(n_1, n_2, n_3, n_4)$ at the output of the interferometer from the measured ones $p_{\mathrm{det}}(n_1, n_2, n_3, n_4)$. The corresponding errors are obtained by error propagation arising from the uncertaintanties introduced with the calibration procedure, and from the statistical Poissonian uncertainty in the probabilities $p_{\mathrm{det}}(n_{1}, n_{2}, n_{3}, n_{4})$.
}

{
\section{Convergence of the Nelder-Mead Algorithm}}

{\subsection{Role of finite statistics}}

\begin{figure}
    \centering
    \includegraphics[width=0.5\linewidth]{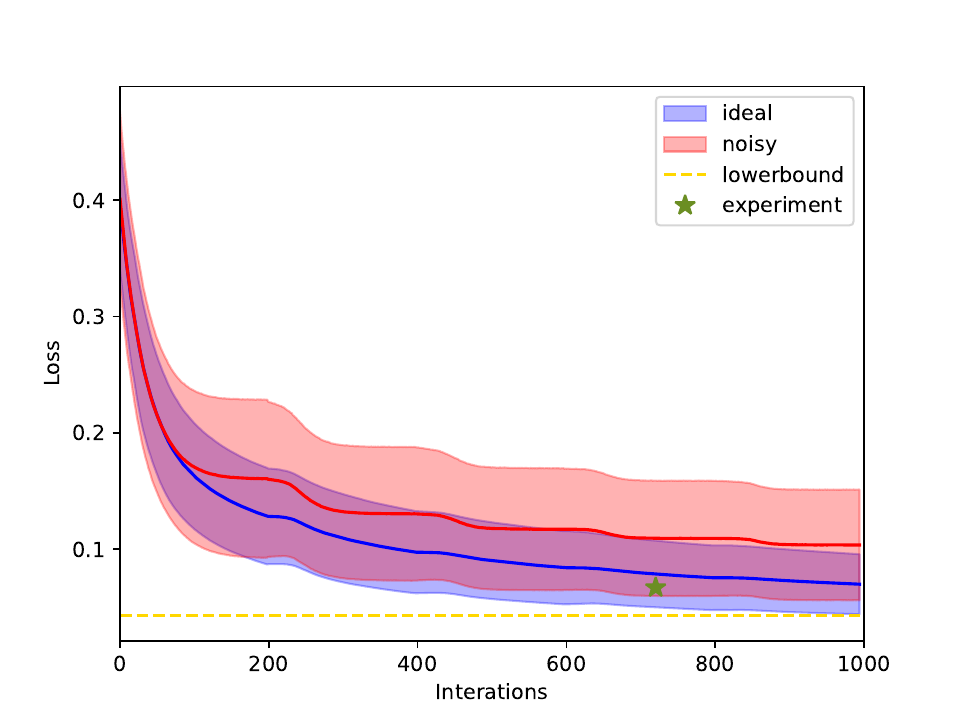}
    \caption{{Average (solid lines) and one standard deviation (shaded area) of the cost function as a function of the Nelder-Mead iterations in the ideal case of infinite statistics (blue) and in the finite sample regime of the experiment (red). The average is measured over $2000$ instances of the algorithm with uniformly random extracted starting points. The experimental point is represented by the green star.}}
    \label{fig:convergence}
\end{figure}

{In this section, we investigate the effect of finite statistics on the convergence of the variational phase-covariant cloning machine. In the experiment, the fidelities with the four target states are estimated from a finite sample of two-fold coincidences recorded at the outputs of the integrated device.  In this regard, we did a simulation aimed at investigating possible issues in the convergence of the Nelder-Mead algorithm that can arise from the estimation of the cost function on the single-photon counts. For this purpose, we simulated 2000 instances of the algorithm by stopping it after 1000 iterations as we did in the experimental implementation.  We considered an ideal regime in which the fidelities and so the cost functions are estimated on an infinite-size sample and then, a finite statistics scenario that reproduces the experimental conditions. The results of the two simulations are reported in Fig. \ref{fig:convergence}. In particular, we report the average trend of the cost function (solid lines) with one standard deviation represented by the shaded area. The experimental value is reported by the star and it is in accordance with the simulation. From such a study, we note that the finite statistics trend is not so distant from the convergence in the ideal case. This indicates that slow convergence is also an intrinsic feature of the Nelder-Mead algorithm when it is applied to a multi-parameter scenario and does not depend significantly on the statistics. Then, we can conclude that the slow convergences and being stuck in local minima are present regardless of the statistics. However, the finite statistics case requires on average more iterations to reach a given value of the cost function.}

{\subsection{Imperfect convergence: asymmetries and phase-dependence in the clones fidelities}}

\begin{figure}[t]
    \centering
    \includegraphics[width=0.5\linewidth]{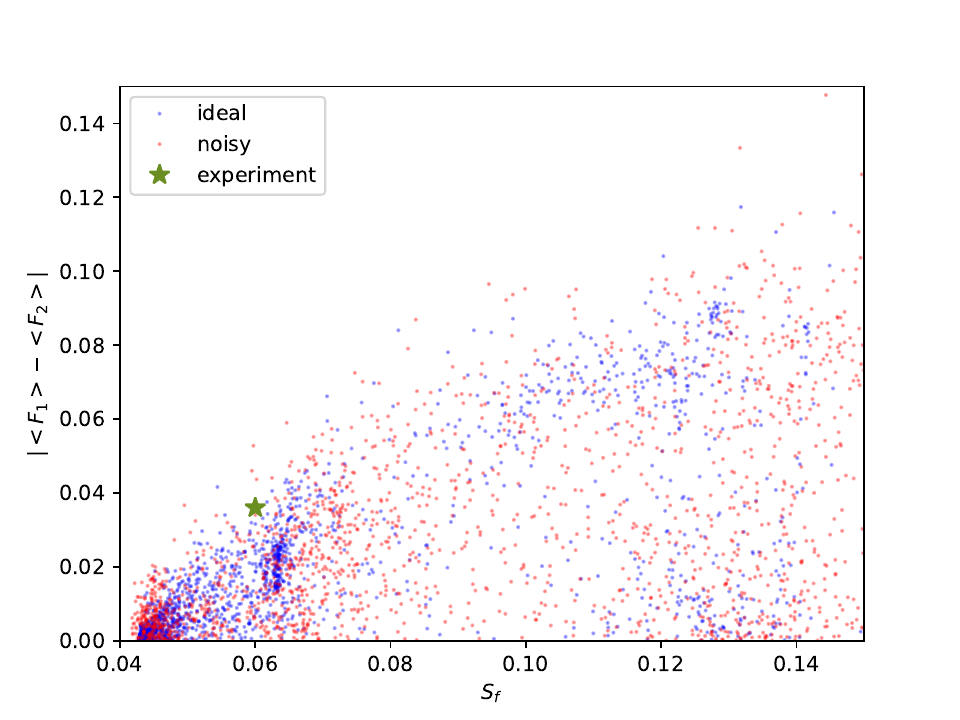}
    \caption{{Residual asymmetry of the phase-covariant variational cloning machine as a function of the final cost function. Blue data are simulations of 2000 runs of the algorithms starting from randomly extracted initial points in ideal conditions, i.e., with infinite statistics. In red, the same simulations with the finite statistics of the experiment. The experimental point is highlighted by the green star}}
    \label{fig:assymmetry}
\end{figure}

{
In the main text, we propose a versatile variational cloner platform by considering the most general optical circuit that operates on two dual-rail qubits. Such an agnostic approach allows for optimizations on different cloning tasks, such as the phase-covariant and the state-dependent, and to be applied in tasks where the solution is unknown a priori. Such a general approach requires optimizing the parameters of a  4-mode universal optical circuit, that, according to the best layout of Ref. \cite{Clements2016}, envisages $12$ variational parameters. The large number of parameters makes searching for the solution more challenging for the Nelder-Mead algorithm and more sensitive to local minima, as we have analyzed in the previous section. Imperfect convergence causes asymmetries in the two clones and a residual state dependence of the cloning performances. In the following, we provide demonstrations of such effects for the phase-covariant cloning machine investigated in the experimental implementation.}

{The Fig.~\ref{fig:assymmetry} shows the asymmetry of the final clones summarized by the difference between the two fidelities $|F_1 -F_2|$ as a function of the final cost function value for the phase-covariant cloning task. In particular, in this simulation, we consider two conditions: the ideal case (blue points) and the case in which the fidelities are computed from single-photon experiments (red points), i.e., from a finite statistic comparable to that of the experiment. As expected from the previous analysis, the finite statistics does not significantly affect the symmetry of the cloners. Both sets of data, the ideal and the finite statistics one, show that if the algorithm does not stop at the actual minimum, there can be residual asymmetry. In particular, the experimental result, highlighted by the star point, is in accordance with such a simulation.}

\begin{figure}
    \centering
    \includegraphics[width=0.5\linewidth]{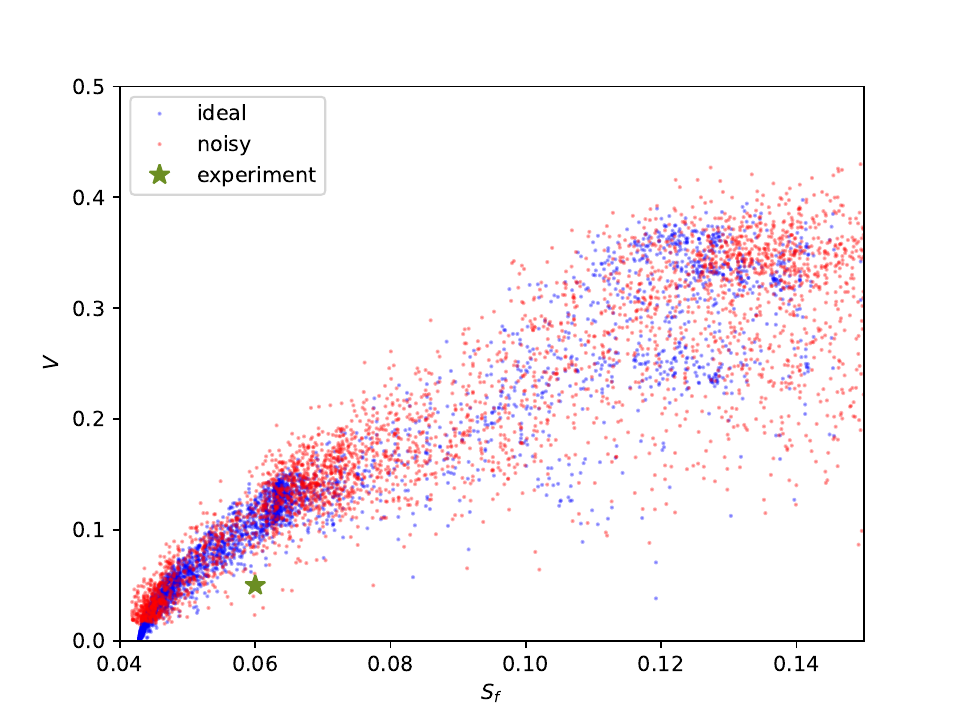}
    \caption{{Fidelity modulation with the phase $\phi$ on the Bloch Sphere. Fringe visibility as a function of the final cost function for the ideal case (blue) and the one with finite statistics (red). The visibility for the two cloners obtained in the experiment is reported by the green star.}}
    \label{fig:visibility}
\end{figure}

{The phase modulation in the two fidelities of the phase-covariant machine is another consequence of an incomplete convergence of the algorithm. To demonstrate this, we performed a further numerical simulation by considering again different final values of the cost function achieved in the optimization. In particular, we studied the visibility of the fringes as a function of the final cost function. The results are reported in Fig.~\ref{fig:visibility}. Also in this case, we can see that the appearance of fringes depends on the convergence of the algorithm. We did the simulation considering the ideal case with infinite statistics (blue points) and the finite statistics of the experiments. 
The fringes that can be observed in Fig. 3e of the main text are compatible with the level of accuracy achieved after 1000 iterations of the Nelder-Mead optimization. The experimental point is reported as the green star in Fig. \ref{fig:visibility}.}

\begin{figure}[ht]
    \centering
    \includegraphics[width=0.8\linewidth]{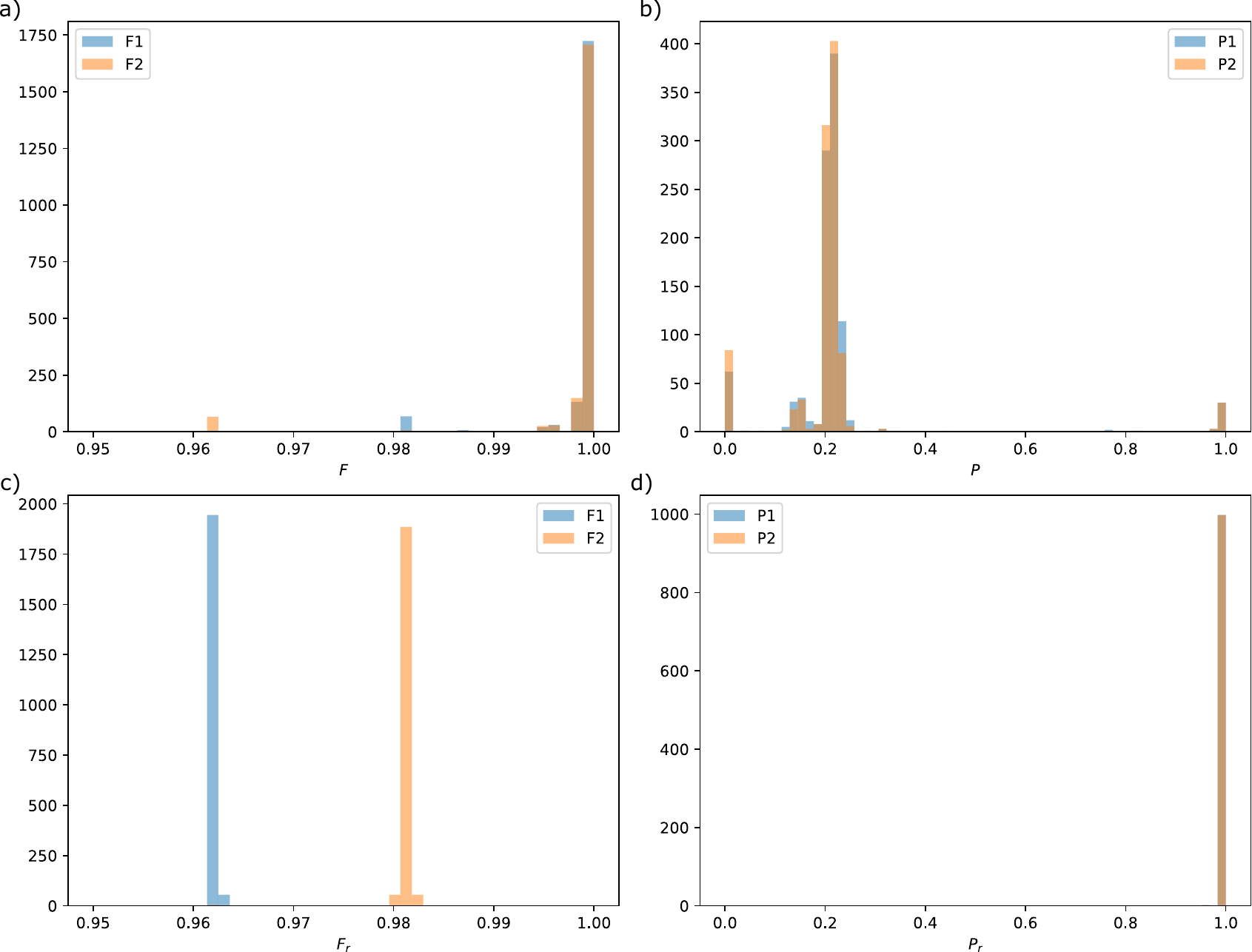}
    \caption{{Effect of the regularization term in the state-dependent variational cloner. In a-b we present the fidelity and the success probability for a couple of states  $(\pi/2,\pi/8)$,$(\pi/2,-\pi/8)$, and $1000$ instances of the minimization with random starting point. In c-d, we present the same result with the regularization term in the cost function.}}
    \label{fig:SD_hist}
\end{figure}

{\section{Photonic state-dependent cloning machine}
In this section, we briefly analyze the variational photonic state-dependent cloning machine and the best design of the cost function. In particular, from the numerical investigation carried out emerges that the cost function requires a regularisation term that takes into account the probability of success. In this way, the variational optimization can solve the optimal state-dependent task by producing pairs of states with the highest achievable fidelity and unit success probability.
To demonstrate such a choice of the cost function, we performed $1000$ optimizations with random starting points both with and without the regularisation term of Eq. 9 of the main text, for the following couple of states $(\pi/2,\pi/8)$,$(\pi/2,-\pi/8)$.
If we perform the optimization without the regularisation term, we obtain the result in Fig.~\ref{fig:SD_hist}a/b.
The first result is that in the $\sim 5\%$ of the instances, the algorithm seems to get stuck in a local minima; in the other cases 
the final fidelity is above $0.99$, however the success probability varies significantly and in $15\%$ of the instances is close to 0 for one of the two states.
If we add the regularization term in the optimization, then the behaviour completely changes (see Fig.~\ref{fig:SD_hist}c/d). In this case, all the instances converge to the minima. The final fidelities decrease slightly and a slight asymmetry appears. However, the main difference can be seen in the success probability, which increases and stabilises at $P = 0.998$.
This shows the effects of the regularization term and its appropriateness to stabilize the minimization process.
}

%